\documentclass[12pt,3p,preprint,sort&compress]{elsarticle}
\pdfoutput=1
\usepackage[colorlinks=true]{hyperref}

\usepackage{amsmath}
\usepackage{amssymb}
\usepackage{mathtools}
\usepackage[utf8]{inputenc}

\newcommand{\e}{{\mathrm{e}}}

\newcommand{\be}{\begin{equation}}
\newcommand{\ee}{\end{equation}}
\newcommand{\ba}{\begin{align}}
\newcommand{\ea}{\end{align}}
\newcommand{\bg}{\begin{gather}}
\newcommand{\eg}{\end{gather}}
\newcommand{\bseq}{\begin{subequations}}
\newcommand{\eseq}{\end{subequations}}

\begin{document}
\begin{flushright}
	INR-TH-2021-012
\end{flushright}

\title{Constraints on light scalars from PS191 results} 
\author[inr,mpti]{Dmitry Gorbunov}
\ead{gorby@ms2.inr.ac.ru}
\author[inr]{Igor Krasnov}
\ead{iv.krasnov@physics.msu.ru}
\author[inr,mpti,lpnhe]{Sergey Suvorov}
\ead{suvorov@inr.ru}
\address[inr]{Institute for Nuclear Research of Russian Academy of Sciences, 117312 Moscow, Russia}
\address[mpti]{Moscow Institute of Physics and Technology, 141700 Dolgoprudny, Russia}
\address[lpnhe]{LPNHE Paris, Sorbonne Universite, Universite Paris Diderot, CNRS/IN2P3, Paris 75252, France}
\begin{abstract}
      We argue that the fixed target experiment PS191 operating on a proton beam of 19.2\,GeV at CERN in the eighties was sensitive to hypothetical light scalars produced by mesons and decaying to charged particles. The experiment was dedicated to searches for sterile neutrinos produced in weak meson decays and decaying into final states with pairs of charged particles: electrons and muons. Two charged tracks from the same point have been adopted as the signal signature. Exploiting the same signature we use the negative results of searches at PS191 and place new limits on the light scalars coupled to the Standard Model (SM) particles via mixing with the Higgs boson. In particular, previously allowed region of masses 100--150\,MeV and mixing above 4$\times 10^{-4}$ is disfavored. Our analysis can be extended straightforwardly to models with other patterns of scalar couplings to SM particles.

\end{abstract}
\date{}

\maketitle

{\bf 1.} 
New physics can be well below the electroweak scale if it contains particles, which interact only feebly with known species forming the Standard Model (SM) of particle physics. This pattern may be preferable from the theoretical side given the gauge hierarchy problem, see e.g.\,\cite{Giudice:2008bi,Giudice:2013yca}, and at the same time can accommodate viable solutions to the phenomenological problems such as neutrino oscillations, dark matter phenomena, baryon asymmetry of the Universe, etc.\,\cite{Shaposhnikov:2008pf,Coy:2021sse}.    Naturally, such models attract special attention as being potentially testable, since direct production of new light particles is kinematically open.      

All the SM particles couple to each other thanks to the SM gauge interactions, which strongly implies that the new particles, if light, must be singlet with respect to the SM gauge group. Hence, the new physics communicates with the known physics through the products of pairs of SM gauge singlets formed by the SM and new fields respectively. The renormalizable interactions, where the coupling constants are dimensionless, are naturally most promising for probing at low energy experiments\,\cite{Batell:2009di,Alexander:2016aln,Beacham:2019nyx}. These interactions are called {\it portals} and in this study, we concentrate on the scalar (or Higgs) portal\,\cite{Patt:2006fw}, which connects a SM singlet scalar field $\phi$ to the SM Higgs doublet $H$ via the following Lagrangian terms 
\begin{equation}
    \label{portal}
    {\cal L}_{SH}=\mu H^\dagger H \phi +\frac{\beta}{2} H^\dagger H \phi^2\,.
\end{equation}
The non-zero vacuum expectation value of the SM Higgs doublet induces mixing between the Higgs boson and the scalar. In the case of small mixing angle $\theta\ll 1$ and light scalar, 1\,MeV\;$<m_\phi<$\;1\,GeV, we are interested in, the lightest mass state in the scalar sector almost coincides with $\phi$. 

The scalar phenomenology is fully determined by its mass $m_\phi$ and mixing angle $\theta$. Indeed, the scalar effectively couples to the SM particles via mixing with virtual Higgs boson, see Refs.\,\cite{Bezrukov:2009yw,Clarke:2013aya} for details. Consequently, the scalar gets produced by the SM particles via this mixing and decays into the SM particles through the same mixing. The light scalar may be a natural part of the extended Higgs sector\,\cite{Khoze:2013uia,Chen:2015vqy,Dev:2017dui}, serve as a messenger to the dark matter\,\cite{Bezrukov:2014nza,Krnjaic:2015mbs,DAgnolo:2018wcn}, play a role of inflaton\,\cite{Bezrukov:2009yw,Bezrukov:2013fca}, etc.         

Searches for a light scalar is a traditional task for experimental particle physics, starting from the seventies when the long hunt for the SM Higgs boson has been initiated\,\cite{Ellis:1975ap}. So far only one fundamental scalar, the Higgs boson, has been discovered\,\cite{Aad:2012tfa,Chatrchyan:2012ufa} and numerous experiments severely constrain the model parameter space, for the most recent bounds see Ref.\,\cite{CortinaGil:2021nts}. However, the strong physical motivation for the light scalars engages new projects, e.g.\,\cite{Alekhin:2015byh,Feng:2017vli,Batell:2019nwo,Batell:2020vqn}, aimed at further investigations in this direction. 

\begin{figure}[!htb]
    \centering
    \includegraphics[width=\textwidth]{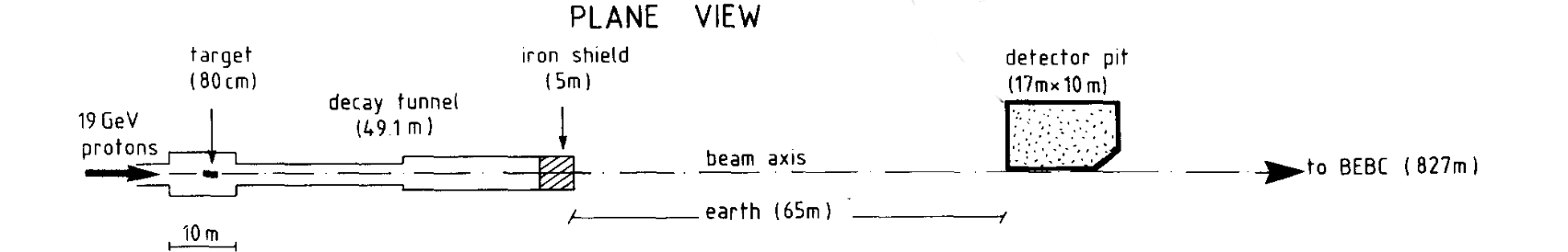}
    \caption{Layout of the PS191 experiment at CERN, adopted from Ref.\,\cite{Bernardi:1985ny}.}
    \label{fig:PS191}
\end{figure}

{\bf 2.} In this letter, we observe that operated in eighties PS191 experiment\,\cite{Bernardi:1985ny,Bernardi:1986hs,Bernardi:1987ek} was capable of searching for the light scalars decaying into charged particles. It was a fixed target experiment, see its layout in Fig.\,\ref{fig:PS191},
with protons of energy 19.2\,GeV hitting a beryllium target (a cylinder 80\,cm long and 6\,mm in diameter) and thus producing mostly light mesons, i.e. pions and kaons. The latter were supposed to source sterile neutrinos: hypothetical fermions, singlet with respect to the SM gauge group, which mix with SM neutrinos and hence appear and decay through the weak interactions. The mixing is small, and if a sterile neutrino were produced by a kaon inside the decay tunnel, see Fig.\,\ref{fig:PS191}, it would travel for a long distance before decay. The decay tunnel was filled with Helium. The cross-section of the decay tunnel was rectangular, with dimensions of 5\,m by 2.8\,m. A 5\,m long iron beam dump was installed at the end of the decay tunnel.

The detector of cross-section  6\,m$\times$3\,m and length $\Delta l=12$\,m was placed at a distance of about $d=128$\,m from the target and $2.29^\circ$ offset from the beam line. It has been designed to measure tracks of charged particles. Sterile neutrinos can decay into two- and three-body states with 2 charged particles providing the signal signature of two charged tracks coming from a single point inside the detector. The downstream side of the detector was occupied by an electromagnetic calorimeter with a hodoscope.  

The experiment PS191 was dedicated to the sterile neutrino searches that have been performed with total statistics of $N_{\text{POT}}=0.86\times 10^{19}$ protons on target (POTs). Its negative results lead to severe bounds\,\cite{Bernardi:1985ny,Bernardi:1987ek}  on sterile neutrino model parameters---masses and mixing with active neutrinos---competitive with present experiments and even future projects, see e.g.\,\cite{Drewes:2018gkc,Kling:2018wct,Krasnov:2019kdc,Gorbunov:2020rjx}.  

\begin{figure}[!htb]
    \centering
    \includegraphics[width=0.9\textwidth]{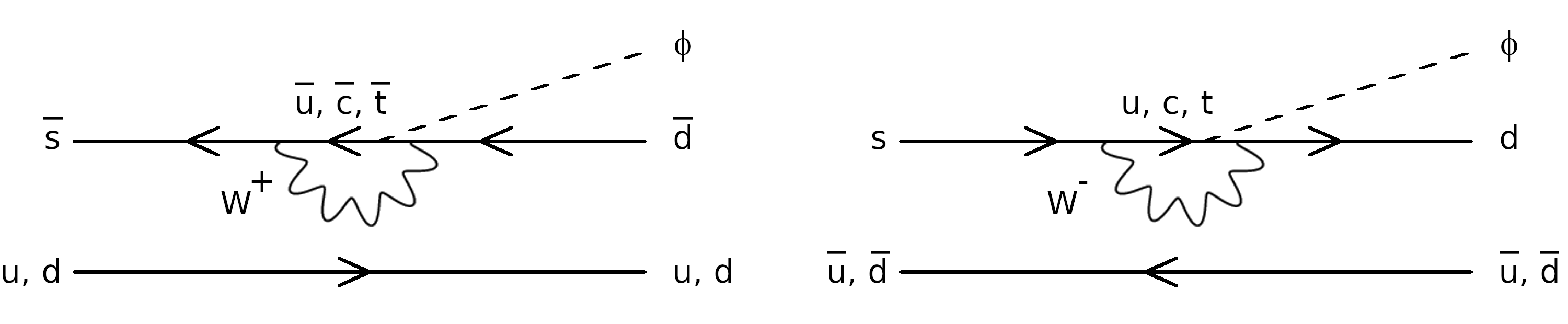}
    \caption{Feynman diagrams for kaon decays into scalar $\phi$.}
    \label{fig:diagram}
\end{figure}

{\bf 3.} The light singlet scalars exhibit very similar to sterile neutrino phenomenology in the framework of PS191. Indeed, they can be produced mostly by kaons in the two-body decay to pions, induced by the one-loop diagram of $s$-$d$ quark transition with virtual $W$-boson and up-quarks inside, see Fig.\,\ref{fig:diagram}. 
The latter emit the scalar via Yukawa-type interactions with virtual Higgs, where the top-quark yields the dominant contribution. Kaon branchings into the scalar are estimated in Refs.\,\cite{Leutwyler:1989xj,Bezrukov:2009yw} as 
\begin{eqnarray}
\text{Br}\left(K^\pm \to \pi^\pm \phi \right) & = & \frac{1}{\Gamma_{K^\pm}}\frac{2p_{\phi_{CM}}}{m_{K^\pm}}\frac{|\mathcal{M}|^2}{16 \pi m_{K^\pm}}\nonumber\\
& = &\frac{9 \tau_{K^\pm}|V_{ts}V_{td}^*|^2 G_F^3 m_t^4 m^2_{K^\pm} p_{\phi_{CM}} \theta^2}{2048\sqrt{2} \pi^5}\nonumber\\
& \approx & 1.6 \times 10^{-3}\;\frac{2p_{\phi_{CM}}}{m_{K^\pm}}\theta^2\,,\\
\text{Br}\left(K_L \to \pi^0 \phi \right) & = & \frac{1}{\Gamma_{K_L}}\frac{2p_{\phi_{CM}}}{m_{K_L}}\frac{|\textrm{Re}\left[\mathcal{M}\right]|^2}{16 \pi m_{K_L}}\nonumber\\
&= & \frac{9 \tau_{K_L}|\textrm{Re}\left[V_{ts}V_{td}^*\right]|^2 G_F^3 m_t^4 m^2_{K_L} p_{\phi_{CM}} \theta^2}{2048\sqrt{2} \pi^5}\nonumber\\
& \approx & 5.7 \times 10^{-3}\;\frac{2p_{\phi_{CM}}}{m_{K_L}}\theta^2\,,
\end{eqnarray}
where scalar 3-momentum in the kaon rest-frame is 
\begin{equation}
p_{\phi_{CM}} =  \frac{M_K}{2}\sqrt{\left(1 - \frac{\left(m_\phi+m_\pi\right)^2}{m_K^2}\right)
\left(1 - \frac{\left(m_\phi-m_\pi\right)^2}{m_K^2}\right)}\,.
\end{equation}

\begin{figure}[!htb]
    \centering
    \includegraphics[width=0.45\textwidth]{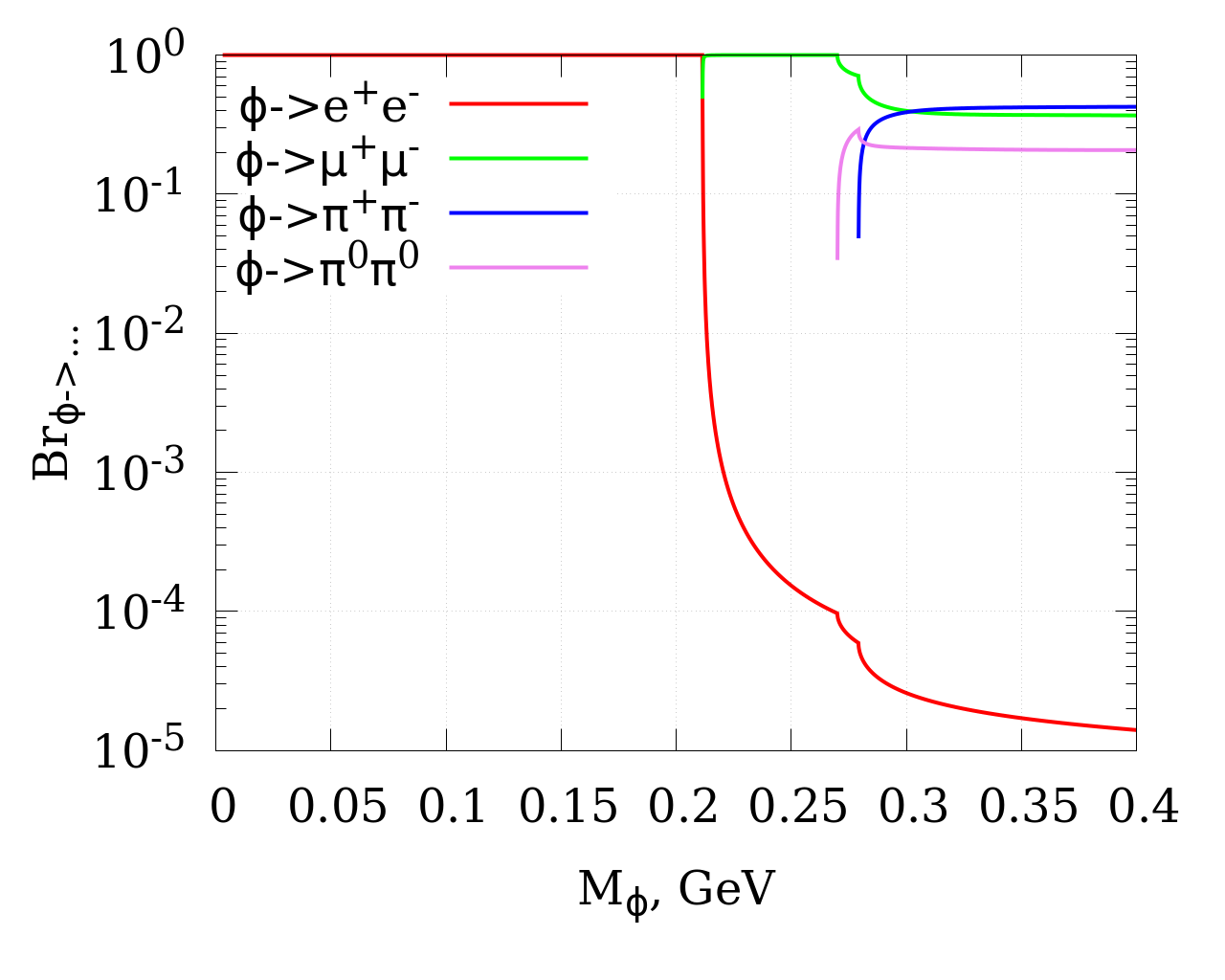}
    \includegraphics[width=0.45\textwidth]{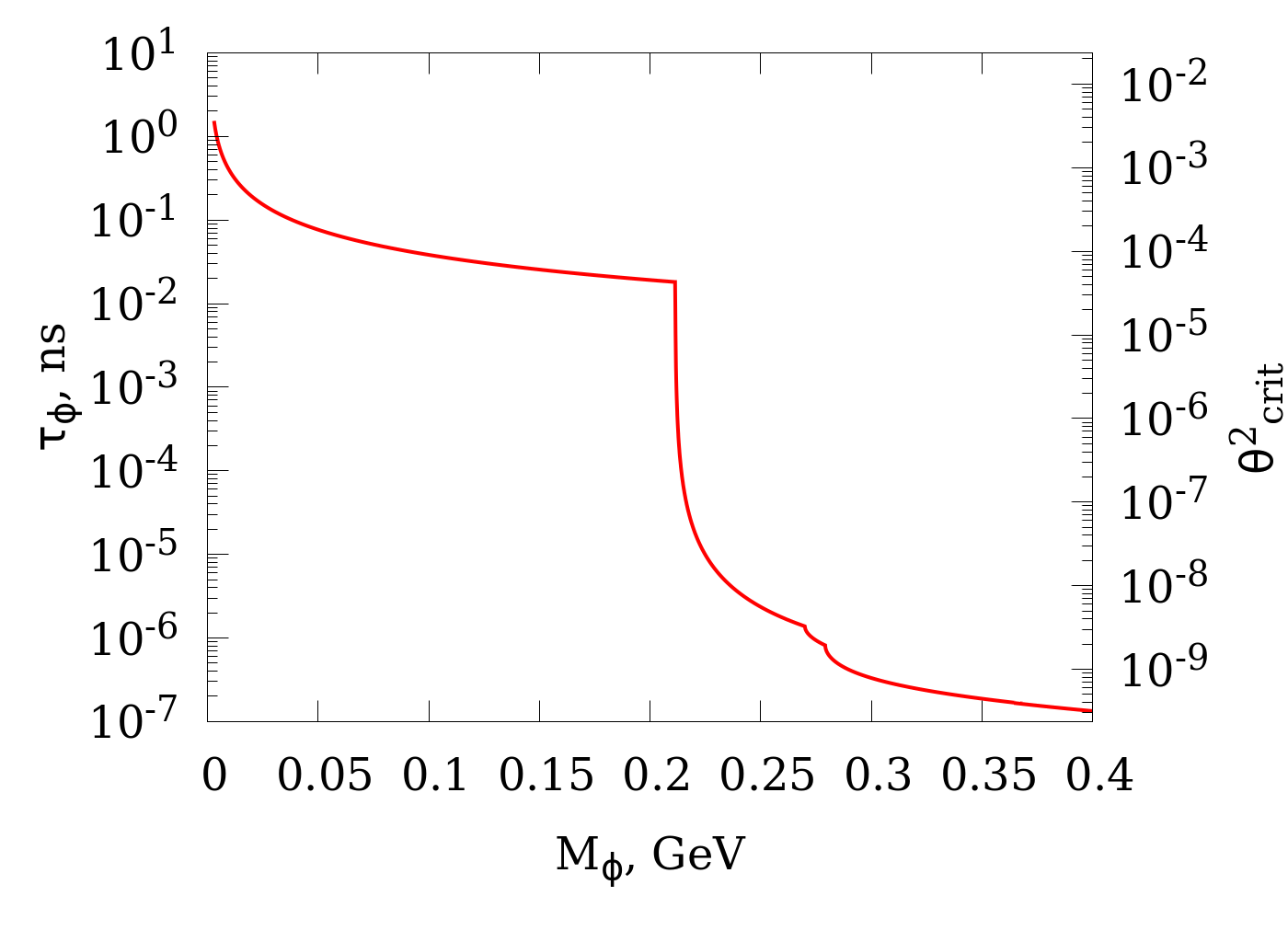}
    \caption{{\it Left panel:} scalar branching ratios. {\it Right panel:} scalar lifetime $\tau_\phi$ for $\theta=1$ and critical mixing $\theta_{\text{crit}}$ (for a weaker mixing the scalar reaches the PS191 detector).}
    \label{fig:branchings+lifetime}
\end{figure}

A part of the scalars then travels towards the detector. If their trajectories pass through the detection area there is a chance to observe the scalar decays. Scalars emerging from kaons can decay only to pions, muons, electrons, and photons. The last channel is strongly suppressed (if the others are kinematically allowed) and gives no charged tracks to be recognizable at PS191. The scalar decay rates to charged lepton  pairs $l^+l^-=e^+e^-,\,\mu^+\mu^-$ and pions read\,\cite{Bezrukov:2009yw} 
\begin{eqnarray}
\label{eq:Gamma1}
\Gamma(\phi \to l^+l^-) &=& \frac{G_Fm_l^2 m_\phi}{4\sqrt{2}\pi} \left(1 - \frac{4m_l^2}{m_\phi^2} \right)^{\!\!\frac{3}{2}} \theta^2\,,\\
\label{eq:Gamma2}
\Gamma(\phi \to \pi^+\pi^-) &=& 2 \Gamma(\phi \to \pi^0\pi^0) =\nonumber\\
&=& \frac{G_Fm^3_\phi}{8\sqrt{2}\pi} \left(\frac{2}{9}+\frac{11}{9}\frac{m_\pi^2}{m_\phi^2}\right)^{\!\!2} \left(1 - \frac{4m_\pi^2}{m_\phi^2} \right)^{\!\!\frac{1}{2}} \theta^2\,.
\end{eqnarray}
The above estimates\,\eqref{eq:Gamma2} for the hadronic modes are robust for the selected mass range of light scalars, but become unreliable for scalar masses about 1\,GeV, see Refs.\,\cite{Donoghue:1990xh,Bezrukov:2009yw,Bezrukov:2018yvd}. The scalar branchings and lifetime for $\theta=1$ are presented in Fig.\,\ref{fig:branchings+lifetime}.  
On the right plot with scalar lifetime, we also outline along the right vertical axis the critical mixing angle $\theta_{\text{crit}}$, below which the scalar of a given mass reaches the detector before decay, $c\tau_\phi>d$. PS191 experiment can probe models with weaker mixing, concentrating on the scalar decay modes into charged particles, $\e^+e^-$, $\mu^+\mu^-$ and $\pi^+\pi^-$, which exhibit the same signal signature as in the case of sterile neutrinos.    

{\bf 4.} To estimate the number of signal events inside the PS191 detector we run Monte Carlo simulations for $N_{sim}=2\times 10^6$ protons of energy 19.2\,GeV. We use GEANT4 toolkit\,\cite{Agostinelli:2002hh}   
to describe the proton scattering off the target material, production of kaons, their propagation inside the target and decay tunnel, and final decay. Based on the study of Ref.\,\cite{Gorbunov:2021jog} we approximate low and high energy QCD processes by BERTini and QGSP models, respectively. 
We account for all the kaons from initiated by protons hadronic showers inside the target, as well as secondaries produced in the decay tunnel, beam dump (iron shield in Fig.\,\ref{fig:PS191}), and in the soil (we utilize sand in the simulations) outside the decay tunnel. In the first interaction, each proton produces about 0.065 $K^+$, 0.032 $K^-$, and 0.044 $K^0_{L}$.  
The total budget of the decayed kaons is given in  Tab.\,\ref{tab:kaon_numbers}. 
\begin{table}[!ht]
    \centering
    \begin{tabular}{c|c|c|c}
    \hline
         &in target and&& in soil outside\\
         &decay tunnel&in the beam dump&decay tunnel \\
         \hline
         $K^+$ produced     & 83833    & 117493 & 133535 \\
         \hline
         $K^-$ produced     & 37439     & 899 & 6984 \\
         \hline
         $K^0_L$ produced   & 19646     & 1824 & 11186 \\
         \hline
         $K^+$ decayed in flight &  92048 & 4572 & 18268 \\
         \hline
         $K^+$ decayed stopped   &  446   & 113028 & 106499 \\
         \hline
         $K^-$ decayed in flight &  40580 & 886 & 3856 \\
         \hline
         $K^0_L$ decayed (in flight) & 25606 & 1571 & 5479 \\
         \hline
    \end{tabular}
    \caption{Decayed kaon budget for a total simulated statistics of $N_{POT}=2\times 10^6$.}
    \label{tab:kaon_numbers} 
\end{table}
In GEANT4 the particle interaction in matter, even e.g., elastic scattering off nuclei, is described as the particle destruction and then production.  In this table, among produced kaons only those which further decay are presented. Namely, kaons that interact with nuclei are omitted from the table. 

\begin{figure}[!ht]
    \centerline{
        \includegraphics[width=0.32\textwidth]{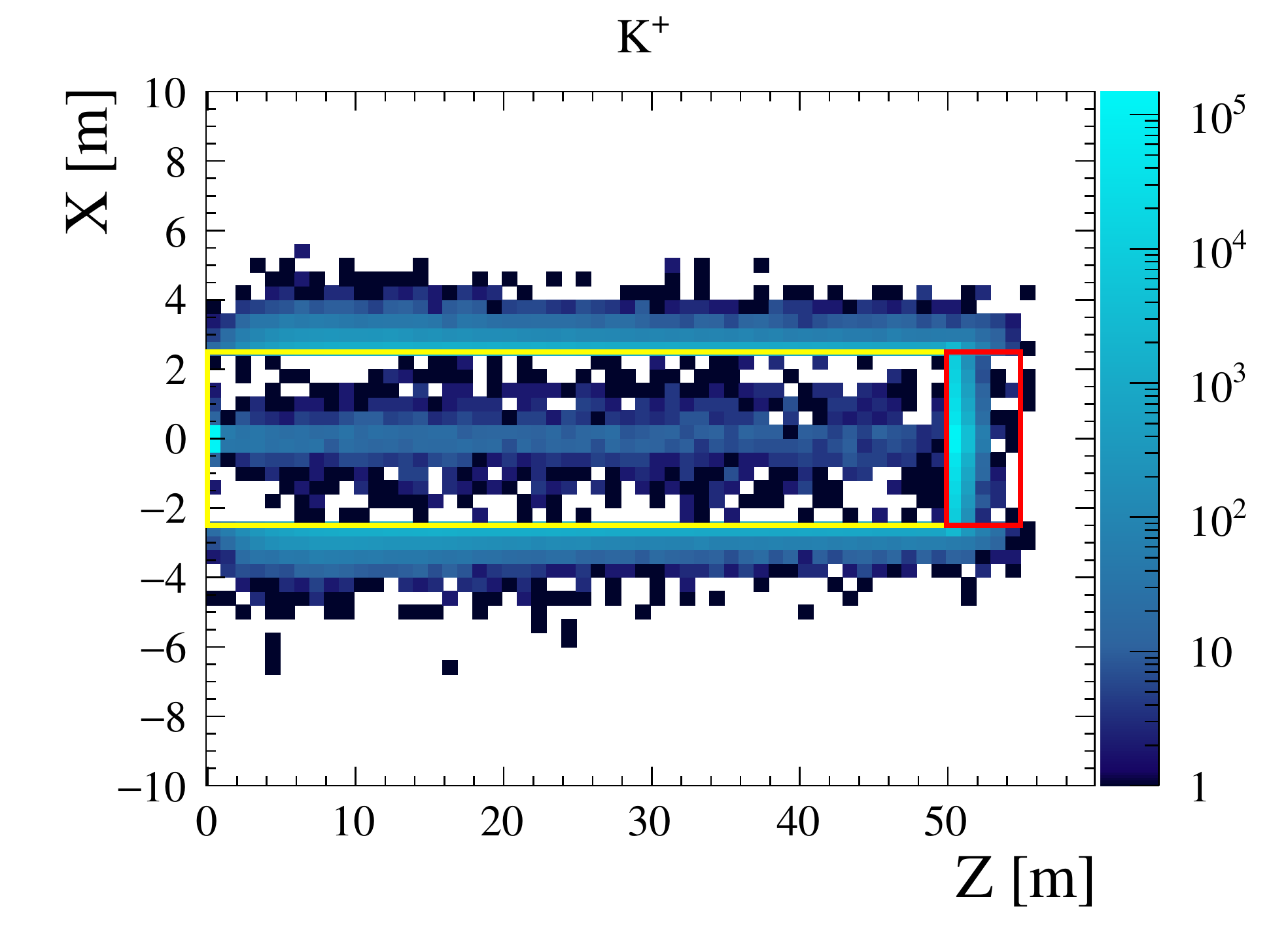}
        \includegraphics[width=0.32\textwidth]{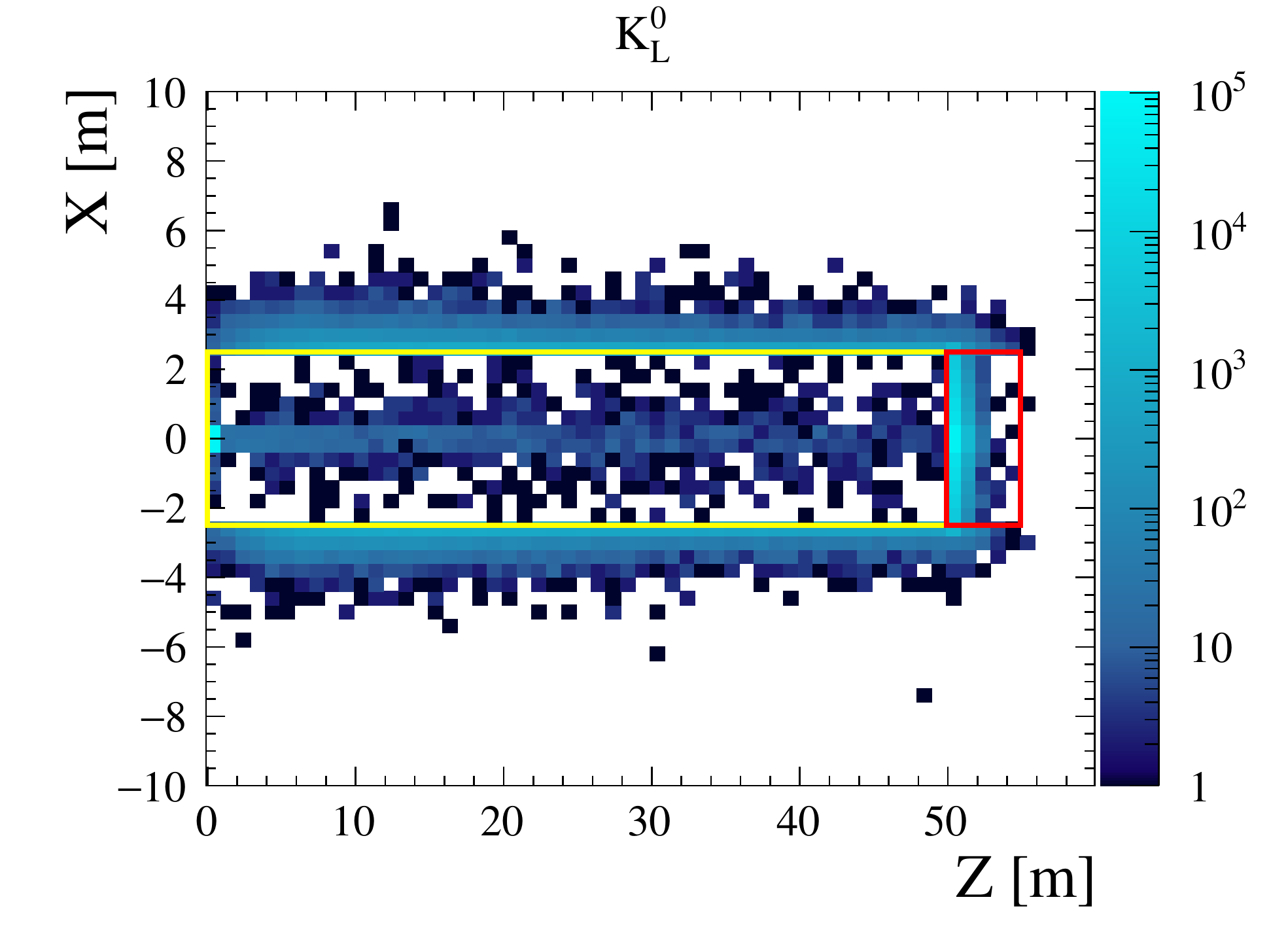}
    }
    \caption{Kaon production points inside the target and decay volume (outlined in yellow), beam dump (outlined in red), and outside (the rest), constrained in the Y direction within the decay volume width $\left|Y\right|<1.4~m$ obtained from the simulation of 19.2\,GeV proton beam collisions with beryllium target. Kaon scattering through the inelastic process is considered as a new kaon production.}
    \label{fig:meson_prod}
\end{figure}

\begin{figure}[!ht]
    \centerline{
        \includegraphics[width=0.32\textwidth]{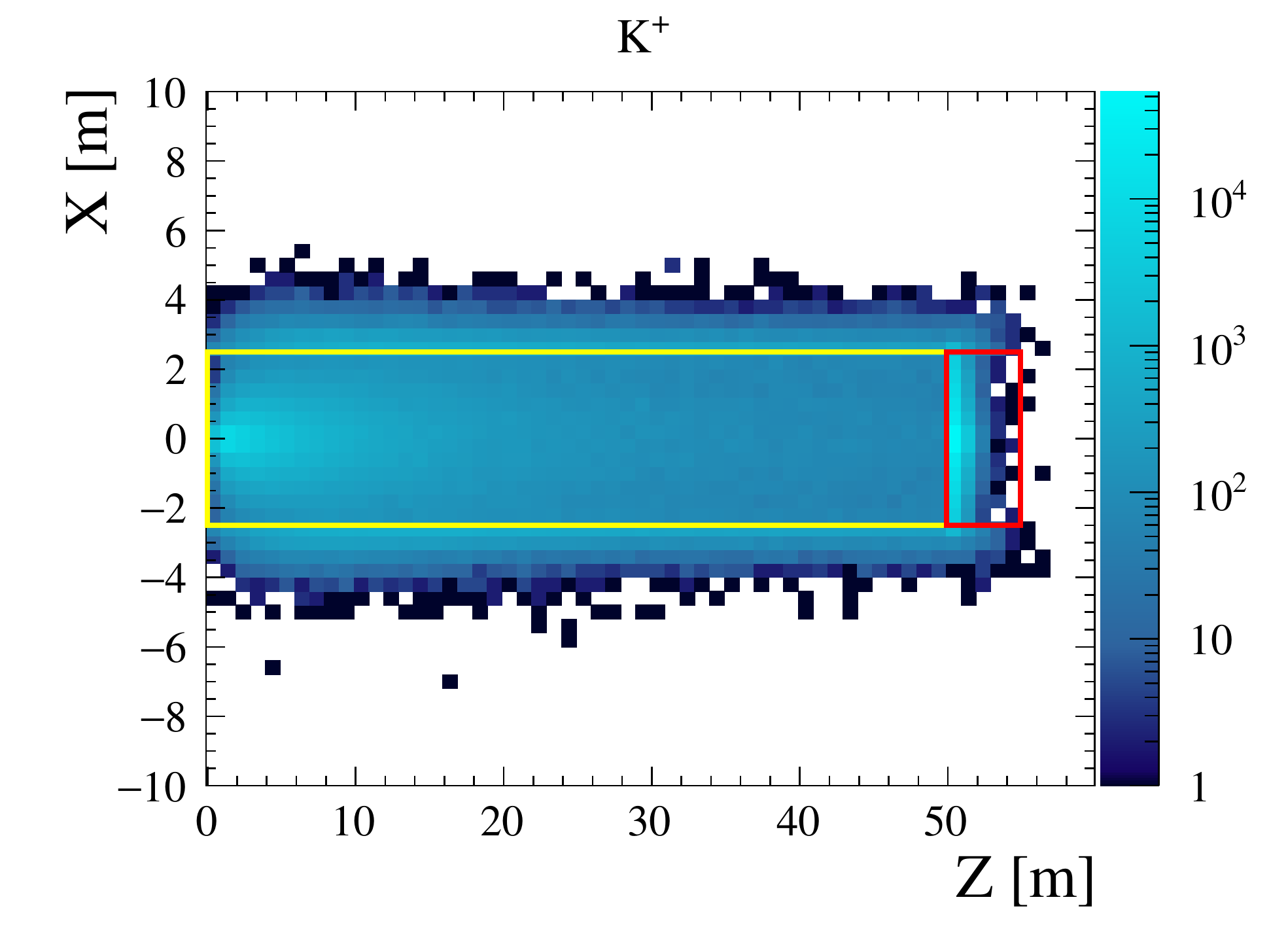}
        \includegraphics[width=0.32\textwidth]{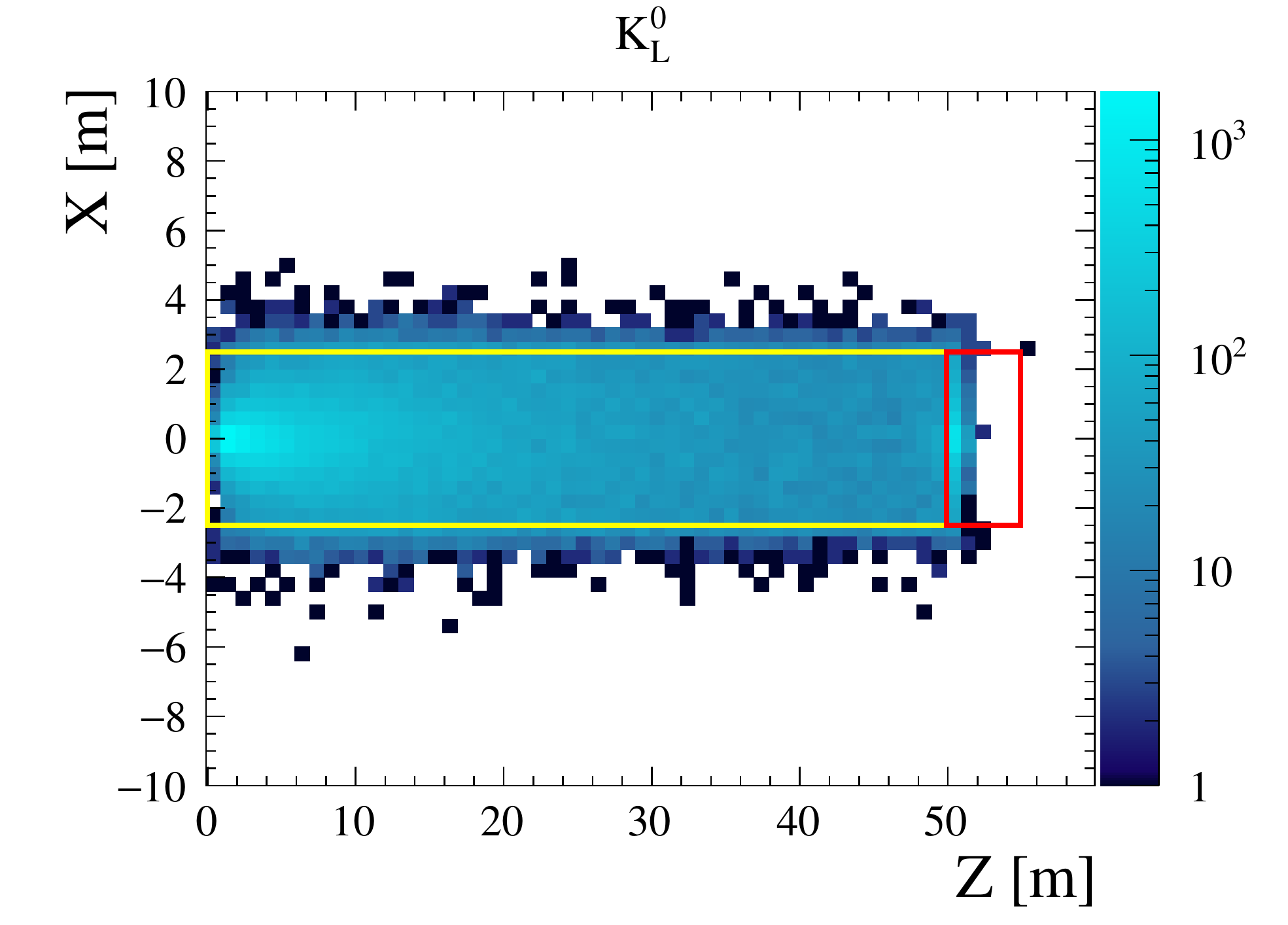}
    }
    \caption{Charged (left panel) and neutral (right panel) kaon decay points inside the target and decay volume (outlined in yellow), beam dump (outlined in red), and outside (the rest) constrained in the Y direction within the decay volume width $\left|Y\right|<1.4$\,m.}
    \label{fig:meson_decay}
\end{figure}

The kaon production and decay position points are shown in Figs.\,\ref{fig:meson_prod}~and~\ref{fig:meson_decay}. 
The main kaon production regions are target and beam dump. Though the kaon outcome from the dump is larger, most of the positively charged kaons decay at rest, thus have a lower probability to provide massive particles going towards the detector. The meson production in the decay volume occurs via the interaction of the beam particles with Helium. Because of the difference in densities, the production there is suppressed with respect to the target, dump, and soil. In the surrounding soil, the kaon production concentrates in the first couple of meters and significantly reduces after that distance. The number of kaons decayed inside the decay tunnel and target somewhat exceeds that of produced there. Here some of the produced kaons exit the decay tunnel before the decay, and some kaons enter the decay tunnel being produced inside the sand, close to the tunnel wall. It happens in our simulations that the income dominates over the outcome. There are no stopped $K^-$ and $K_L^0$ decays since these kaons are captured by the nuclei.

\begin{figure}[!ht]
    \centerline{
        \includegraphics[width=0.32\textwidth]{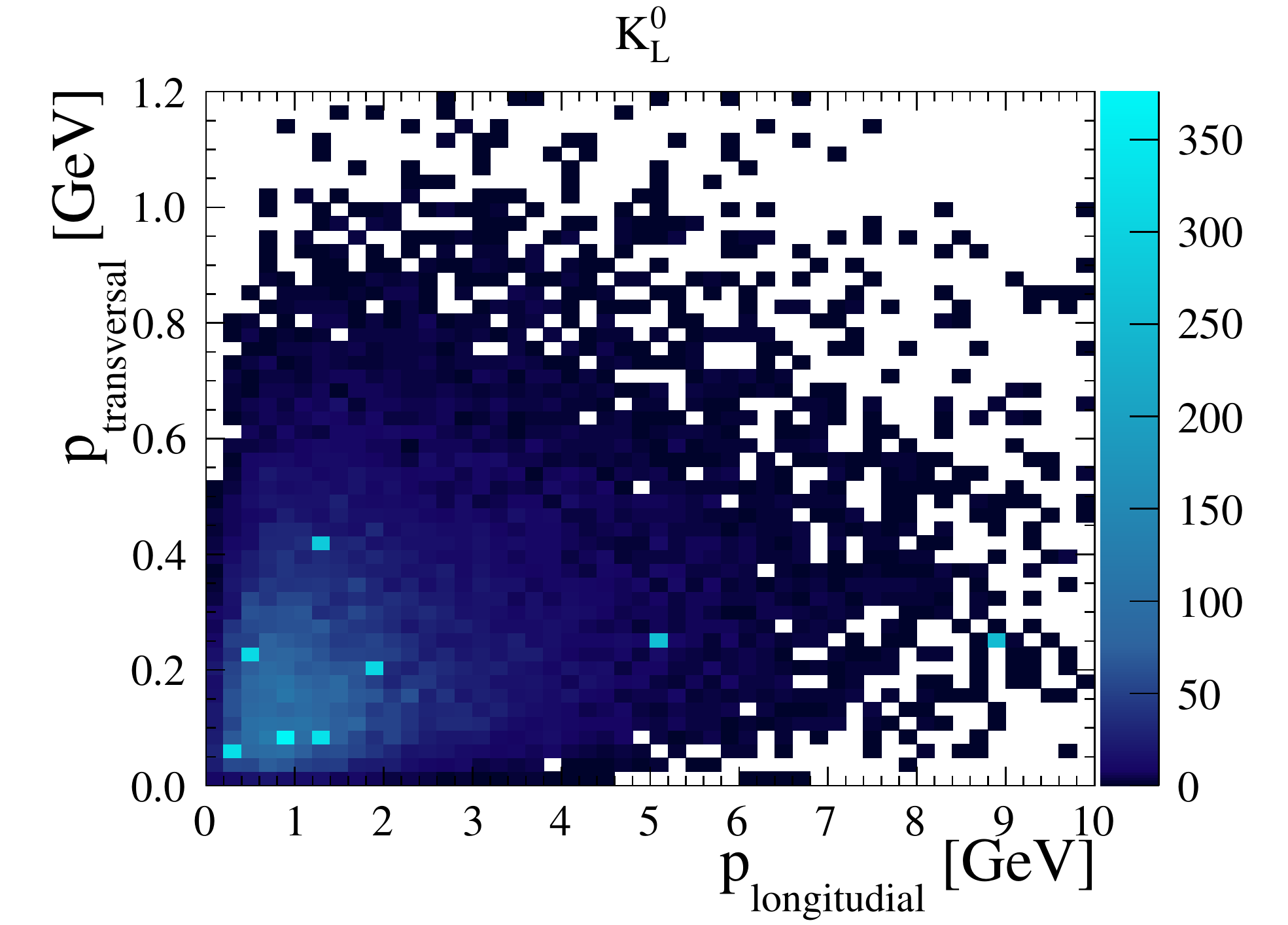}
        \includegraphics[width=0.32\textwidth]{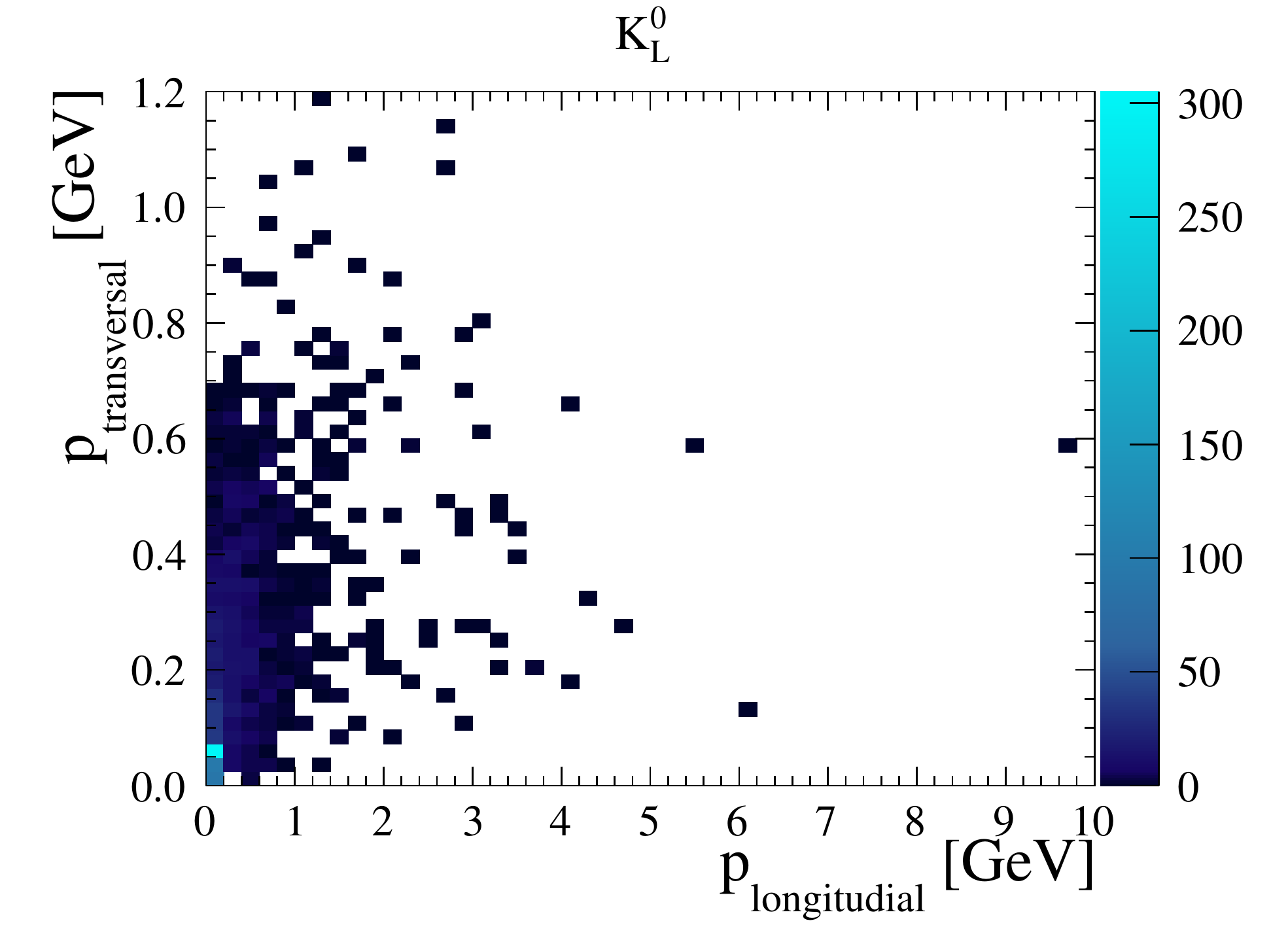}
        \includegraphics[width=0.32\textwidth]{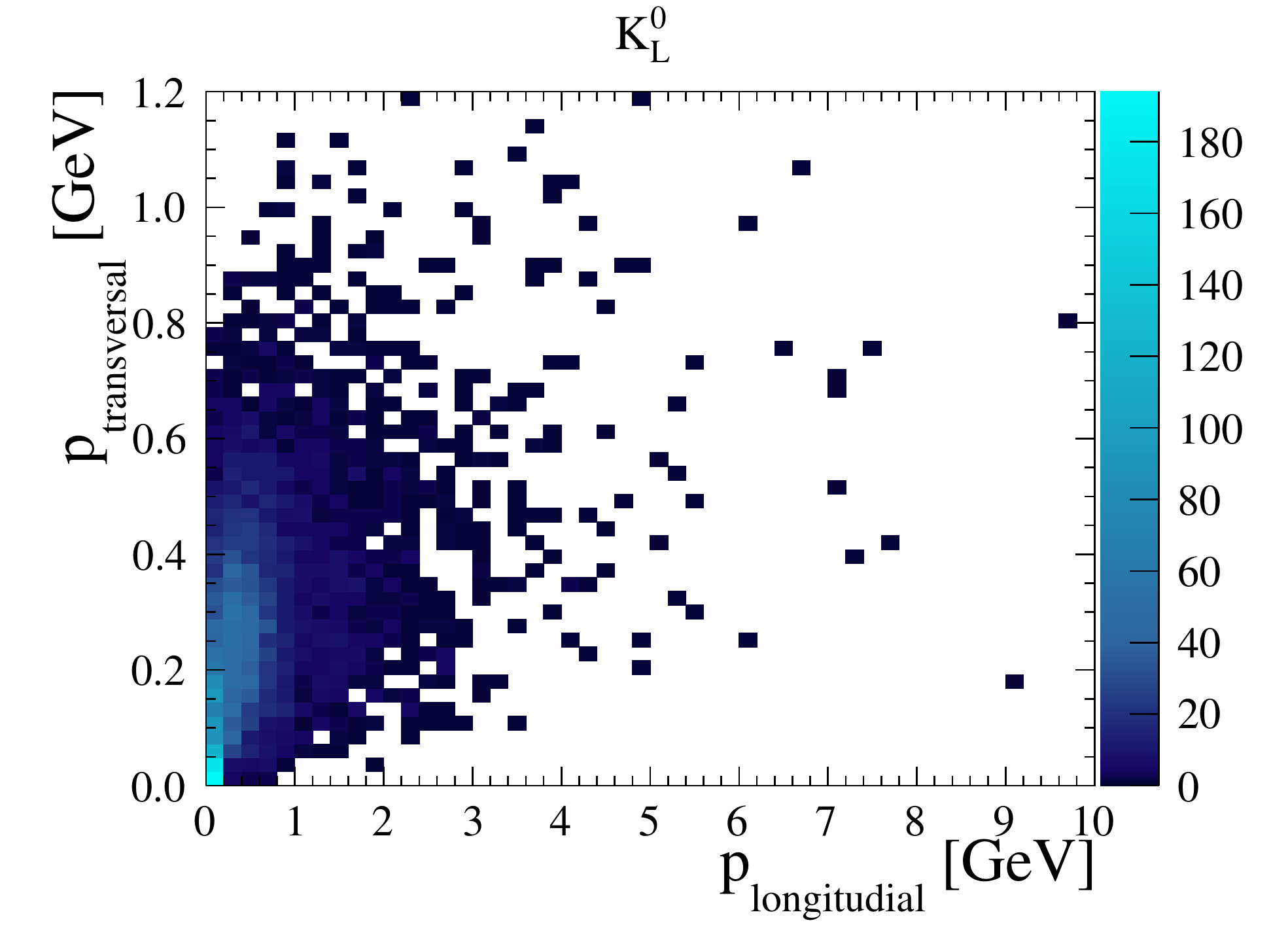}      
    }
     
        \centerline{
        \includegraphics[width=0.32\textwidth]{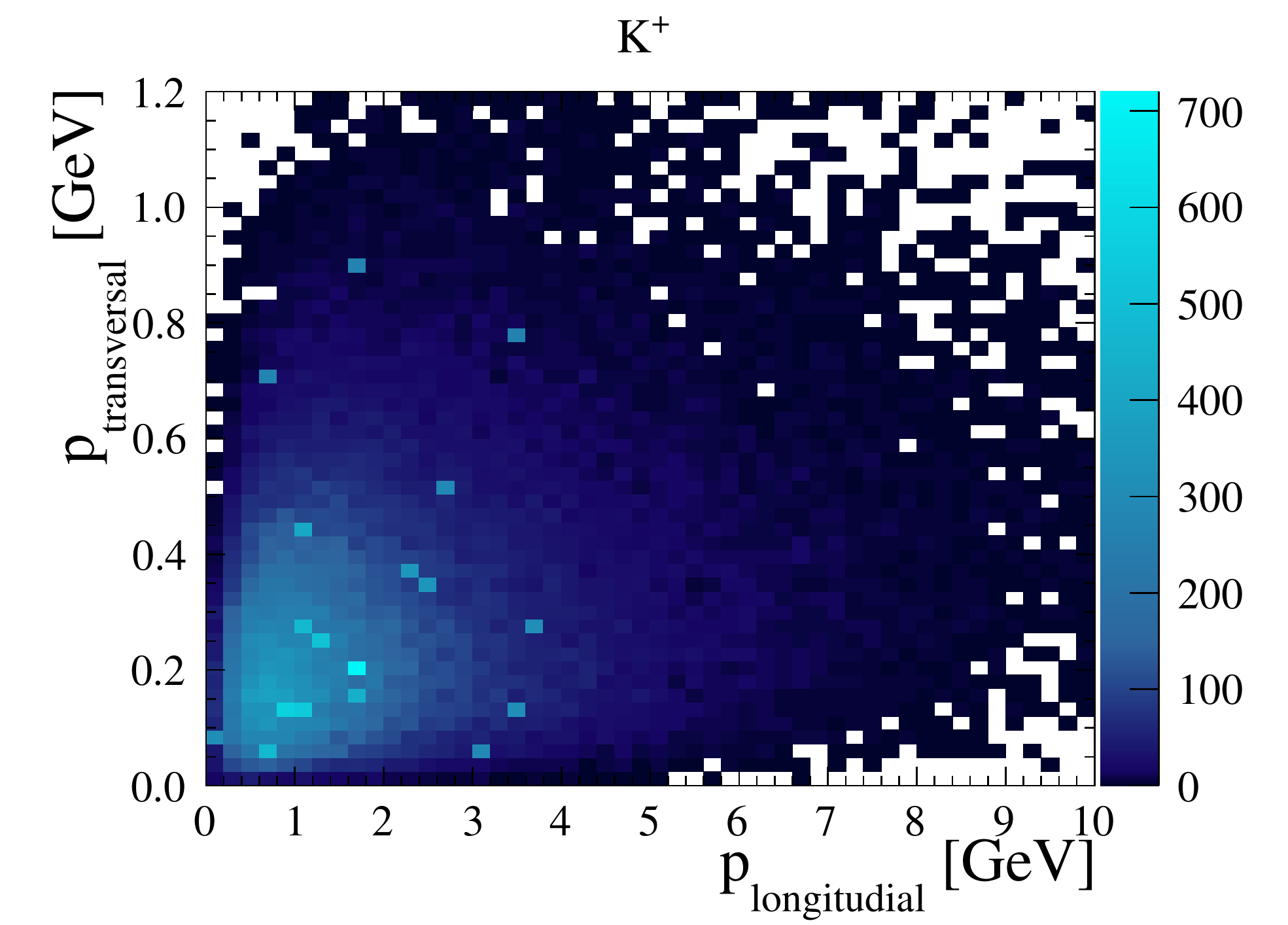}
        \includegraphics[width=0.32\textwidth]{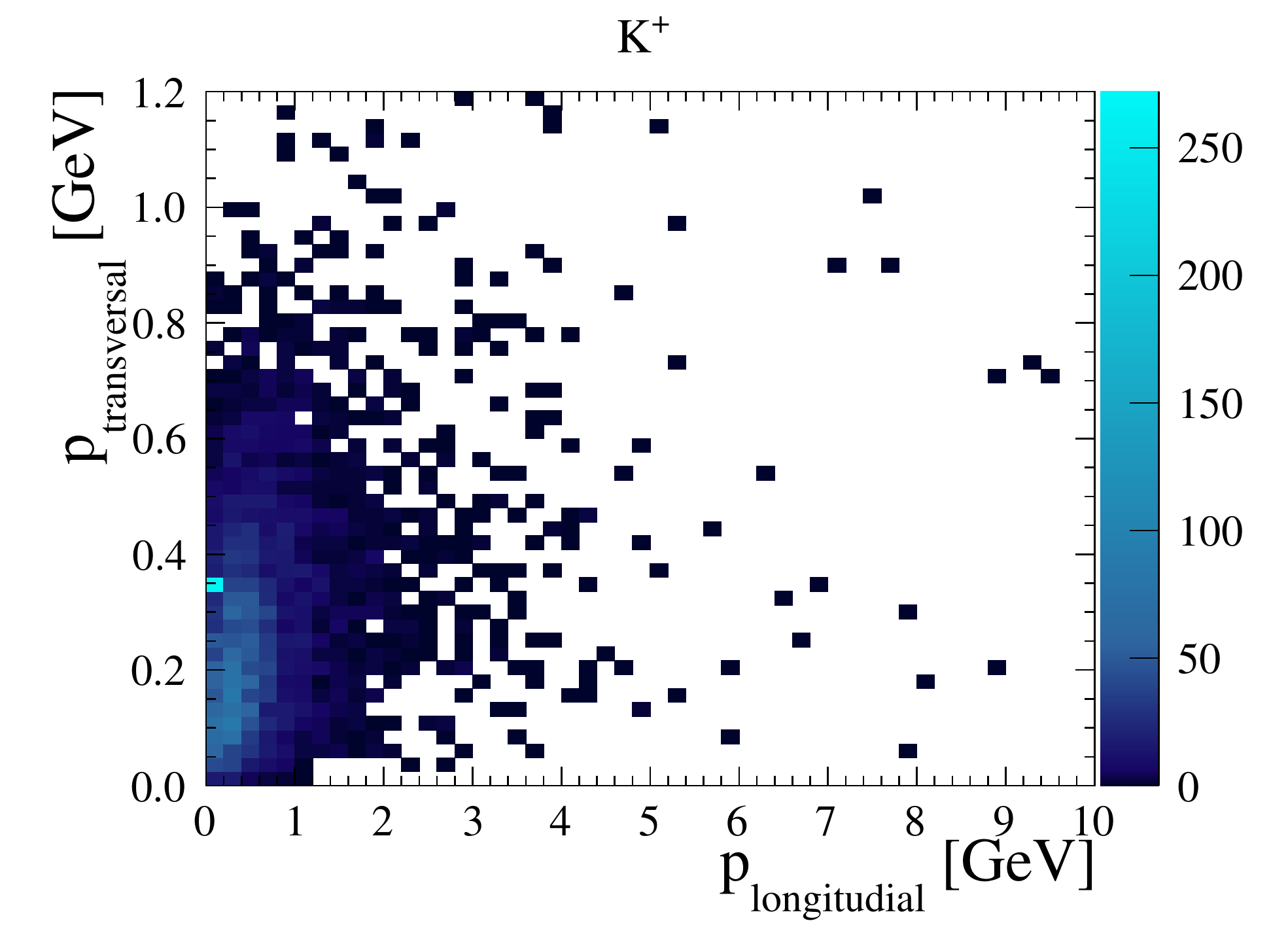}
        \includegraphics[width=0.32\textwidth]{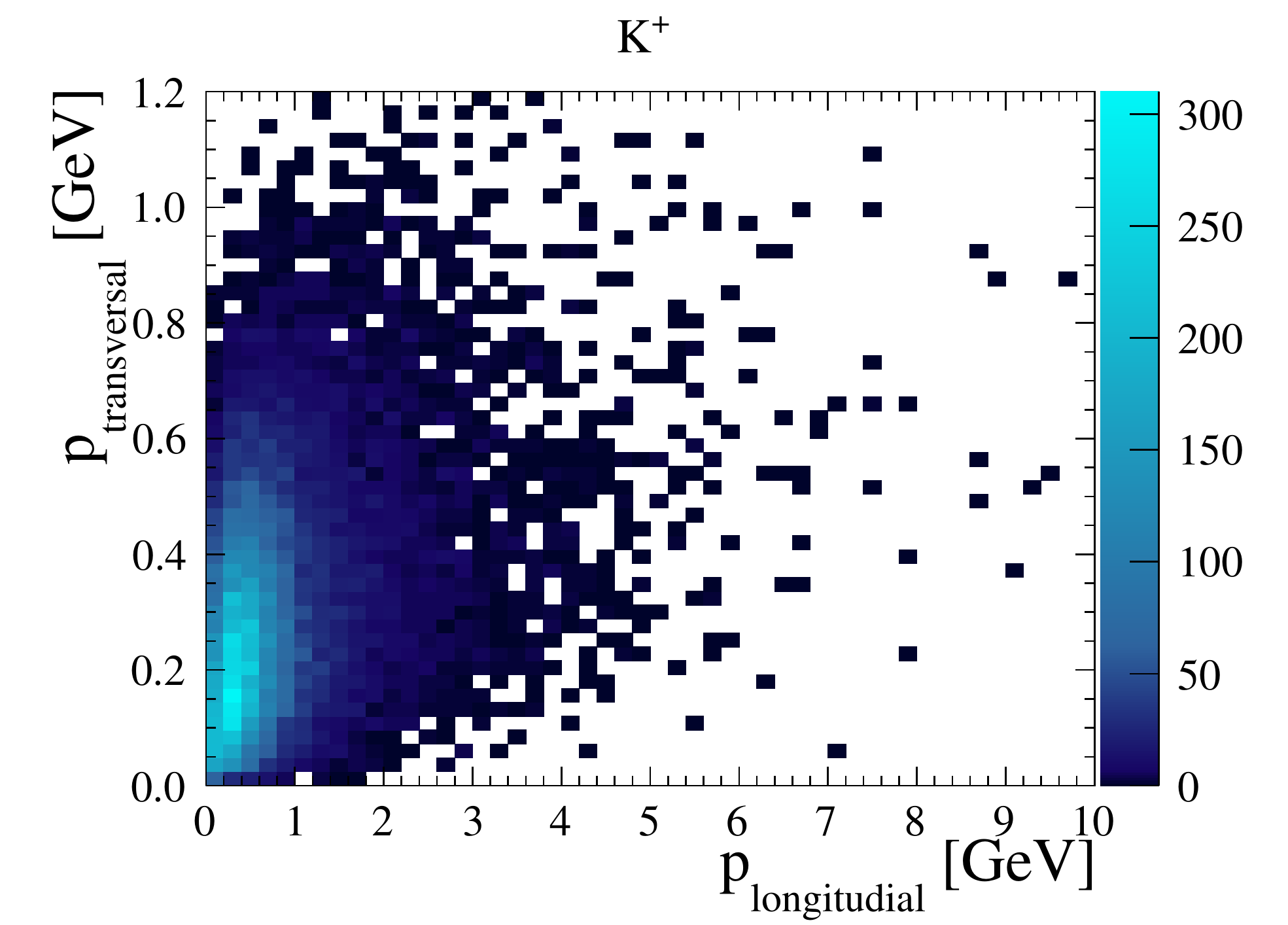}
    }
    \caption{Kinematic distributions of the neutral (top plots) and charged (bottom plots) kaons decayed in the decay volume, beam dump, and sand (from left to right).}
    \label{fig:meson_kinem}
\end{figure}

The kaon kinematic distributions at the moment of the decay are presented in Fig.\,\ref{fig:meson_kinem}.

In PS191 experiment there was a special magnet device to focus positively charged mesons,  while the neutral mesons keep a straight line on the travels.
The device operated only part-time, see Ref.\,\cite{Bernardi:1985ny} for details, corresponding to statistics of $0.56\times10^{19}$\,POTs. The rest, $0.30\times 10^{19}$\,POTs developed without any artificial magnetic field.  

Naturally, the focusing should enhance the ability of PS191 in testing the models with light scalars, since positively charged particles (and $K^+$ fraction is significantly higher than $K^-$, see Tab.\,\ref{tab:kaon_numbers}) produced in the target are stronger aligned with the beam axis in that case. Therefore, first, scalars from $K^+$ have a narrower angular spread and higher chances to cross the detector. Second, more charged particles reach the dump (and less hit the decay tunnel walls), hence generating more kaons there. The budget in Tab.\,\ref{tab:kaon_numbers} implies that the majority of kaons decaying in the beam dump have been produced there. The same is true for kaons decaying outside the decay tunnel. Then, Fig.\,\ref{fig:meson_kinem} shows that kaons decaying inside the beam dump and outside the decay tunnel have noticeably lower energy than those decaying inside the decay tunnel, and so they widely distributed in flying direction. We expect this spread is not wider in the focusing regime, and we expect more kaons in the dump and fewer kaons outside the decay tunnel. Then, the dump is closer to the detector, and hence in the focusing mode, the overall geometry is more promising for catching scalars inside the detector. However, simulation of the kaon production with operating magnets requires detailed knowledge of the PS191 facility. Instead, we use the simulations without any artificial magnetic field and use the whole statistics of $N_{POT}=0.86\times 10^{19}$\,POTs for {\it neutral and positive kaons}, and statistics of $N_{POT}=0.3\times 10^{19}$\,POTs for {\it negative kaons}, thus underestimating the signal flux. The constraints in the scalar model we infer below from this study are conservative. They can be refined with proper simulations including the magnet system installed in PS191.

\begin{figure}[!htb]
    \centerline{
    \includegraphics[width=0.32\textwidth]{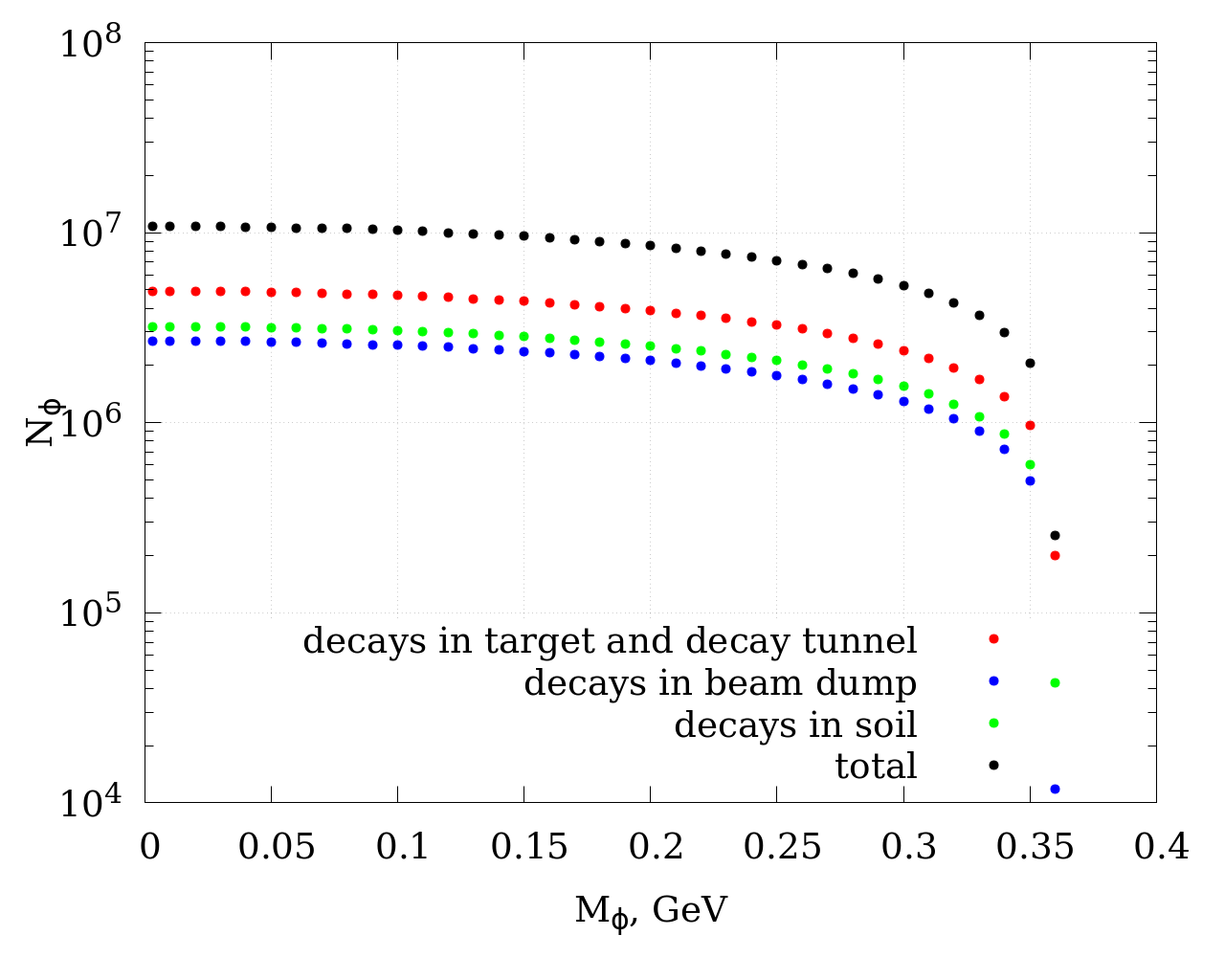}
    \includegraphics[width=0.32\textwidth]{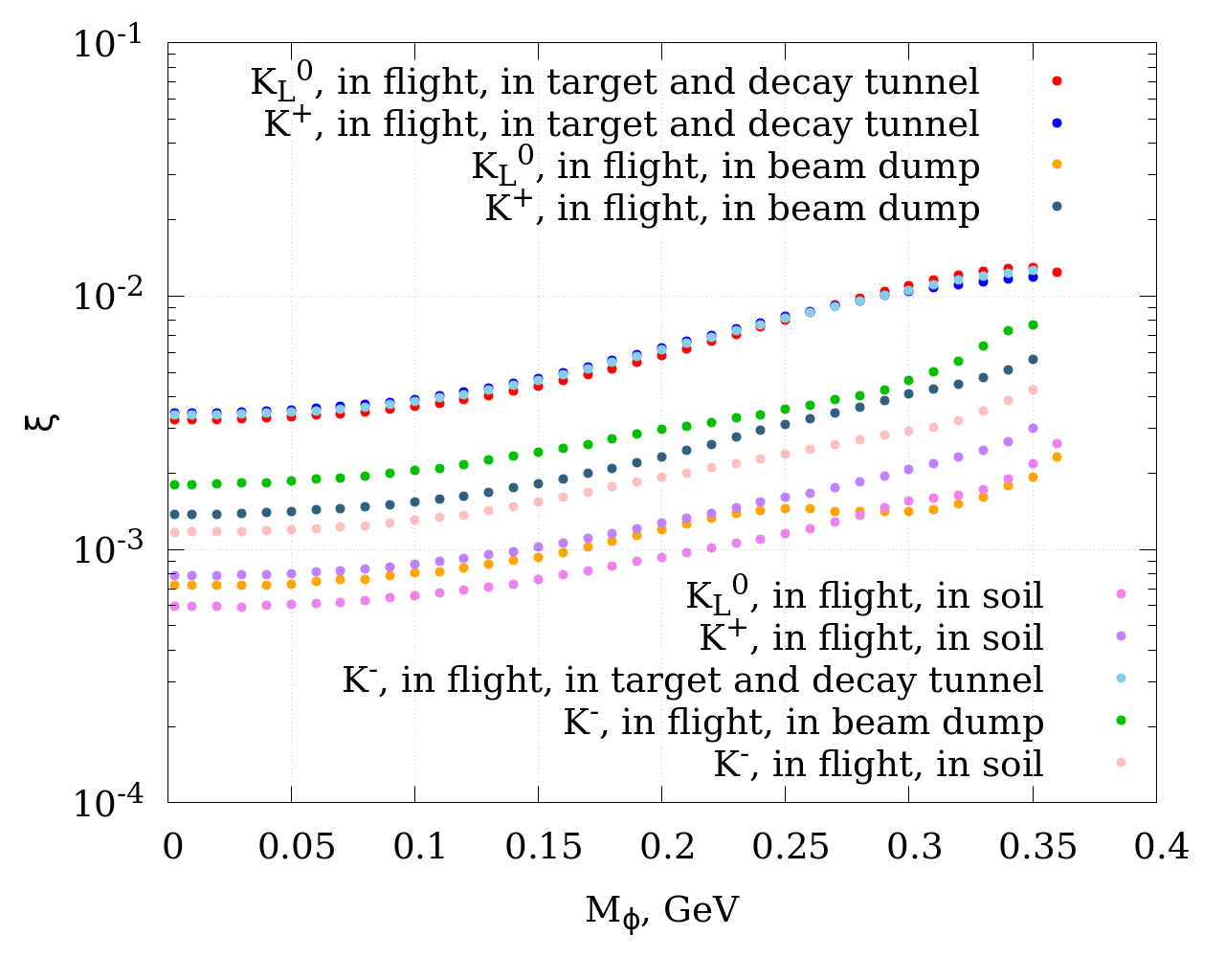}
    \includegraphics[width=0.32\textwidth]{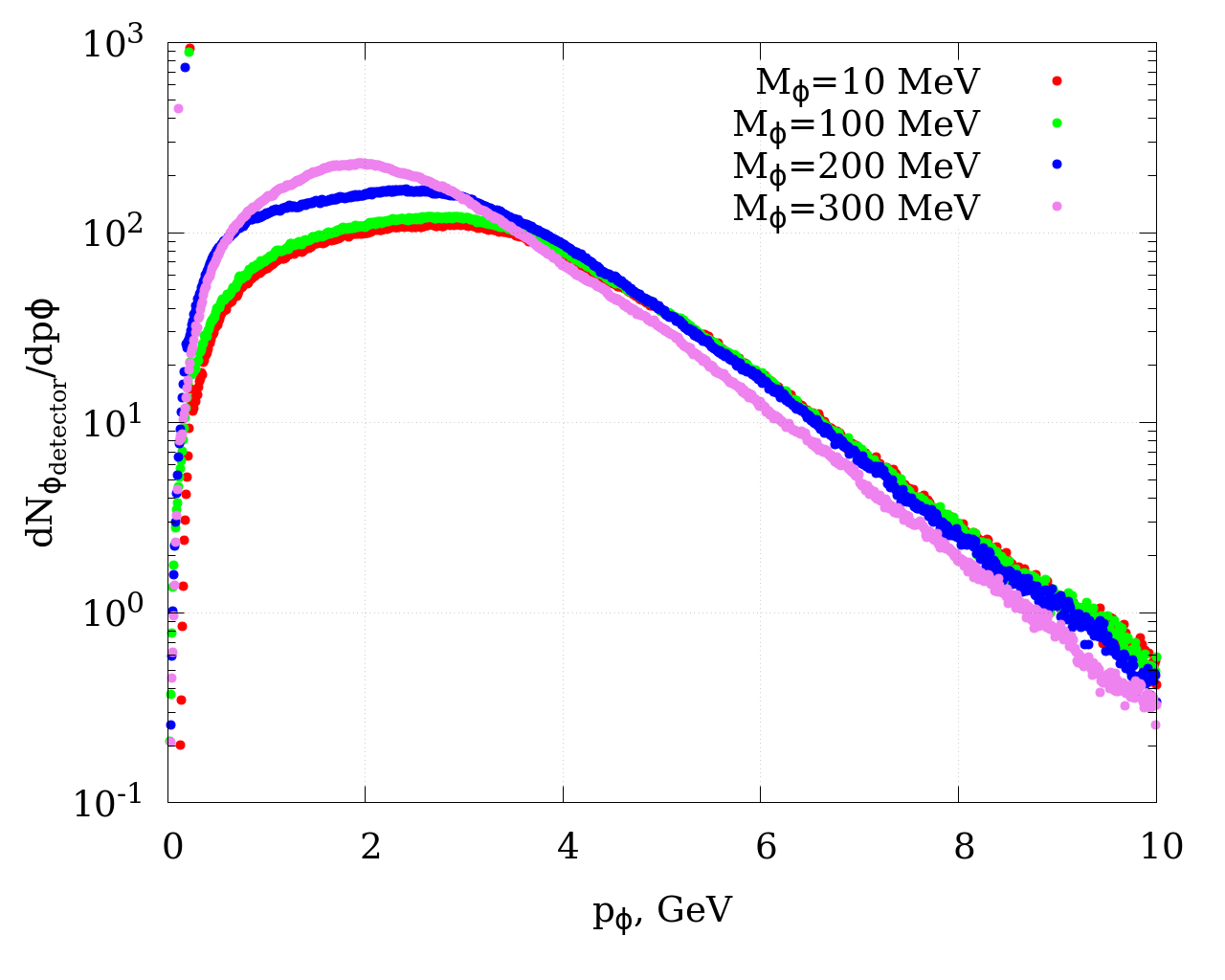}}
    \caption{{\it Left panel:} a total number of the light scalars expected to be produced in kaon decays for $\theta^2=10^{-8}$. {\it Middle panel:} the geometrical factor $\xi$ for the light scalars as a function of its mass for light scalars produced in neutral and charged kaon decays. {\it Right panel:} spectra of the light scalars that reach the detector for a set of masses and $\theta^2=10^{-8}$.}
    \label{fig:xi+spectrum}
\end{figure}

{\bf 5.} 
Light scalar production points obviously coincide with kaon decay points. The total number of produced scalars is presented in Fig.\,\ref{fig:xi+spectrum}. 
We see, that the scalar production is dominated by events from the decay tunnel, which justifies our ignoring  the magnetic field for positive and negative kaons and supports the statement, that the resulting limits we obtain are conservative. 

\begin{figure}[!ht]
    \centerline{
\includegraphics[width=0.32\textwidth]{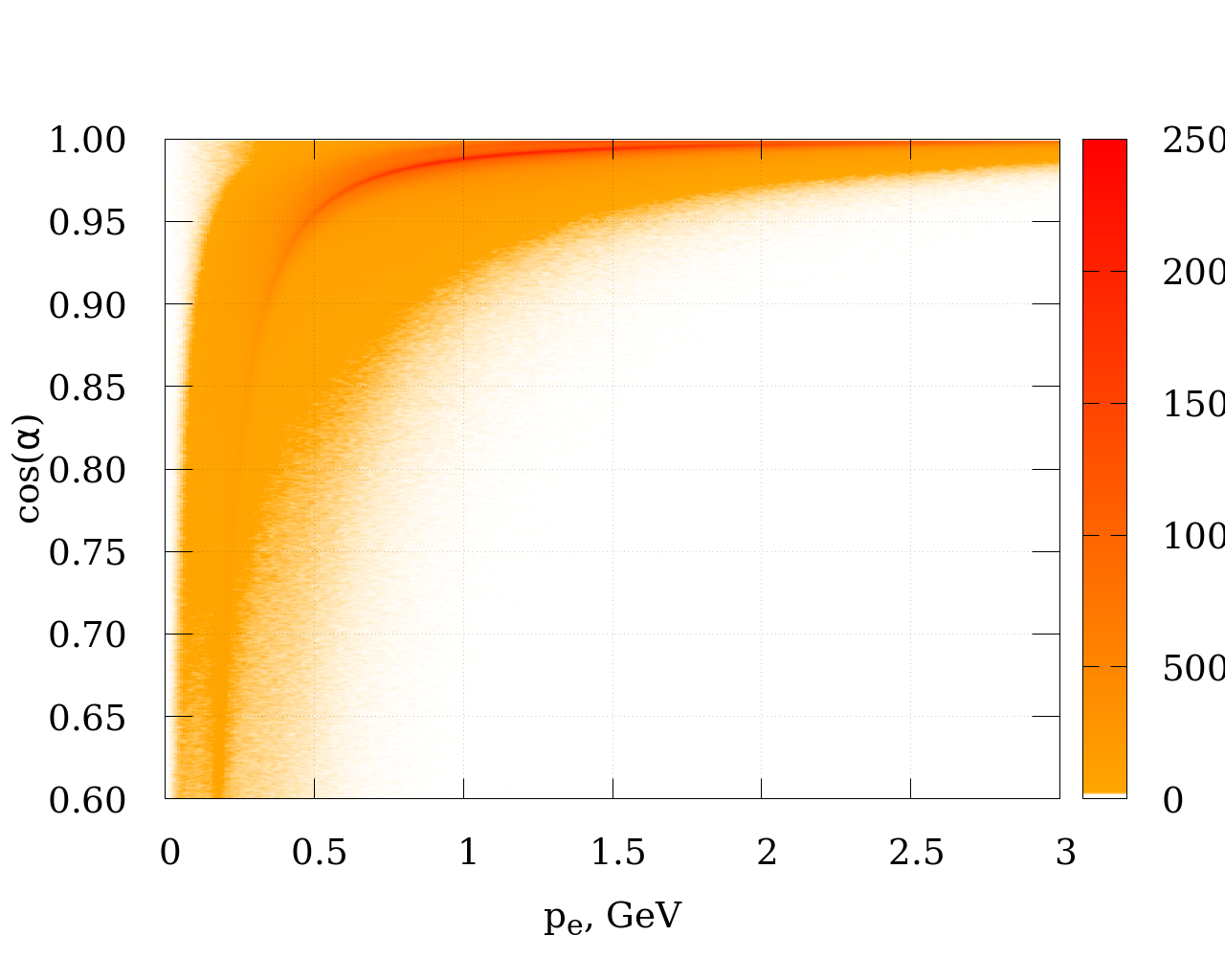}
\includegraphics[width=0.32\textwidth]{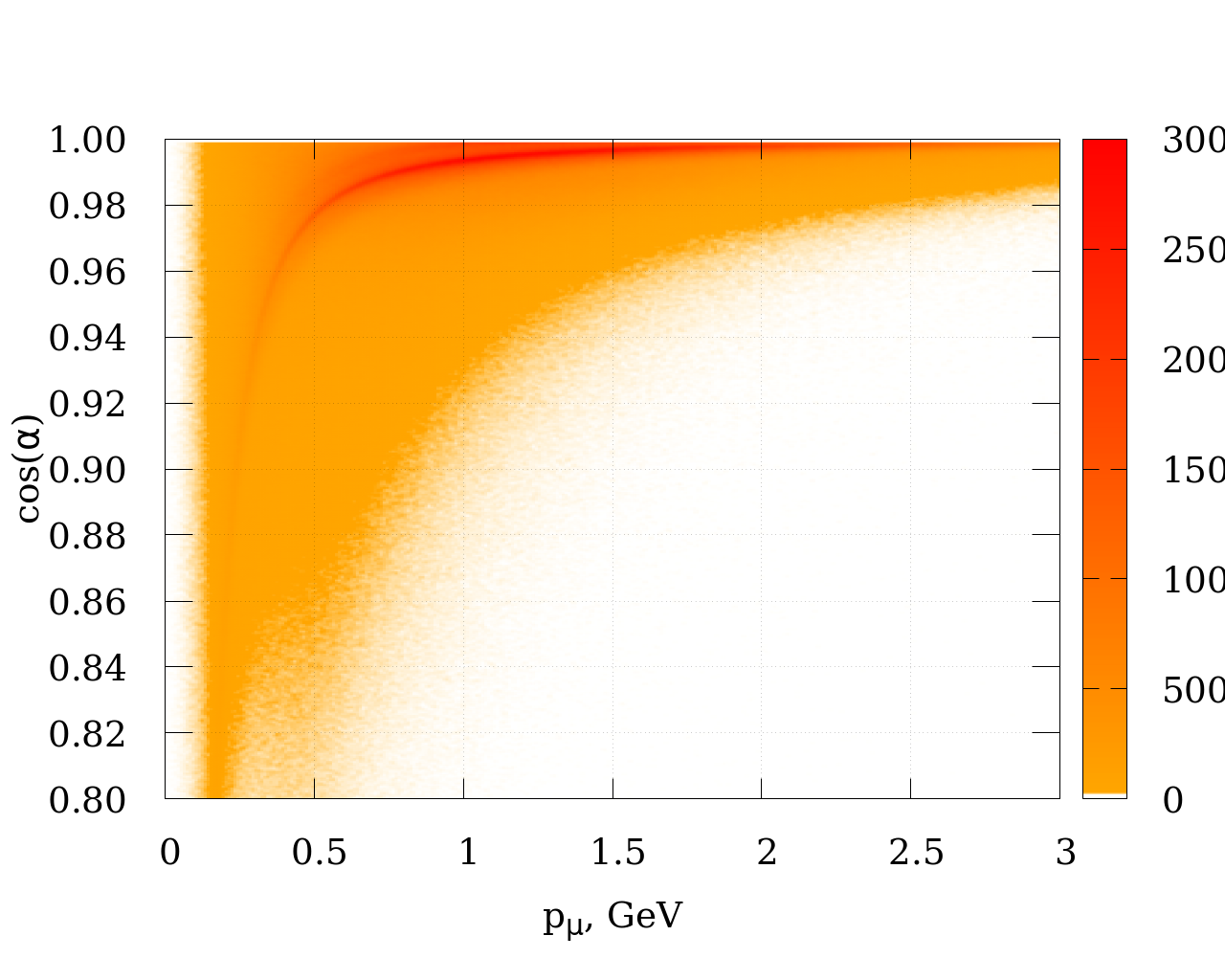} 
\includegraphics[width=0.32\textwidth]{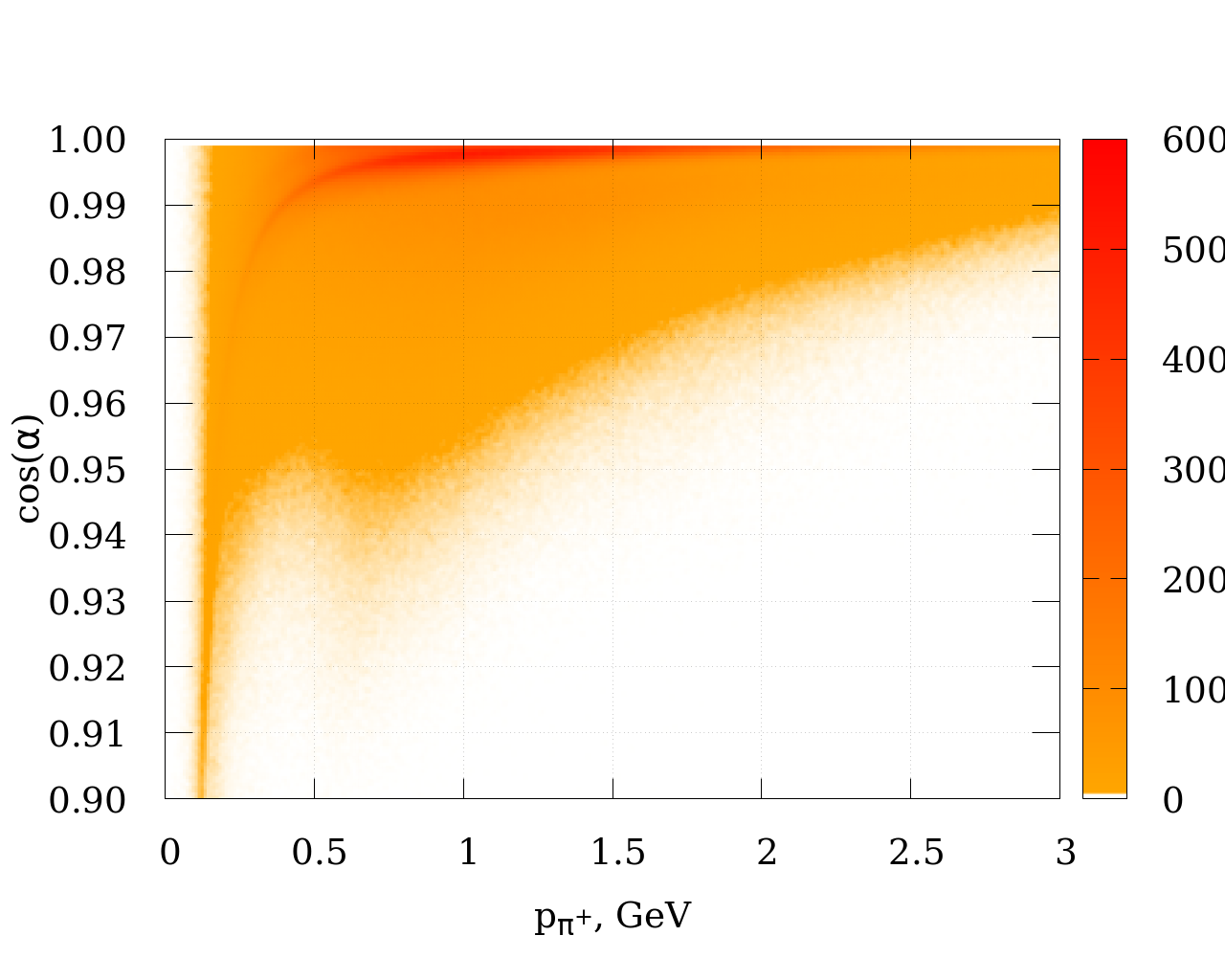}
}
        \centerline{
        \includegraphics[width=0.32\textwidth]{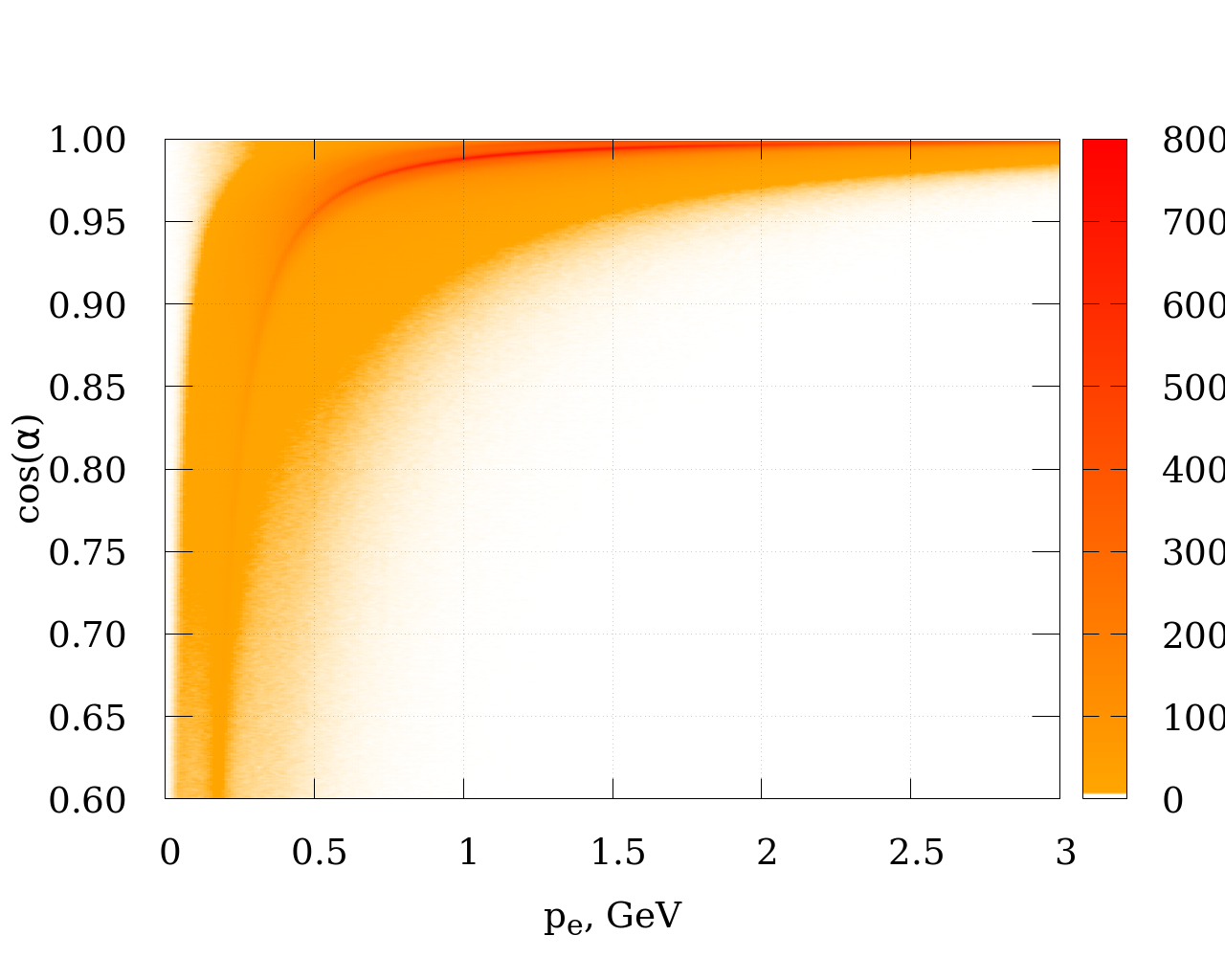} 
        \includegraphics[width=0.32\textwidth]{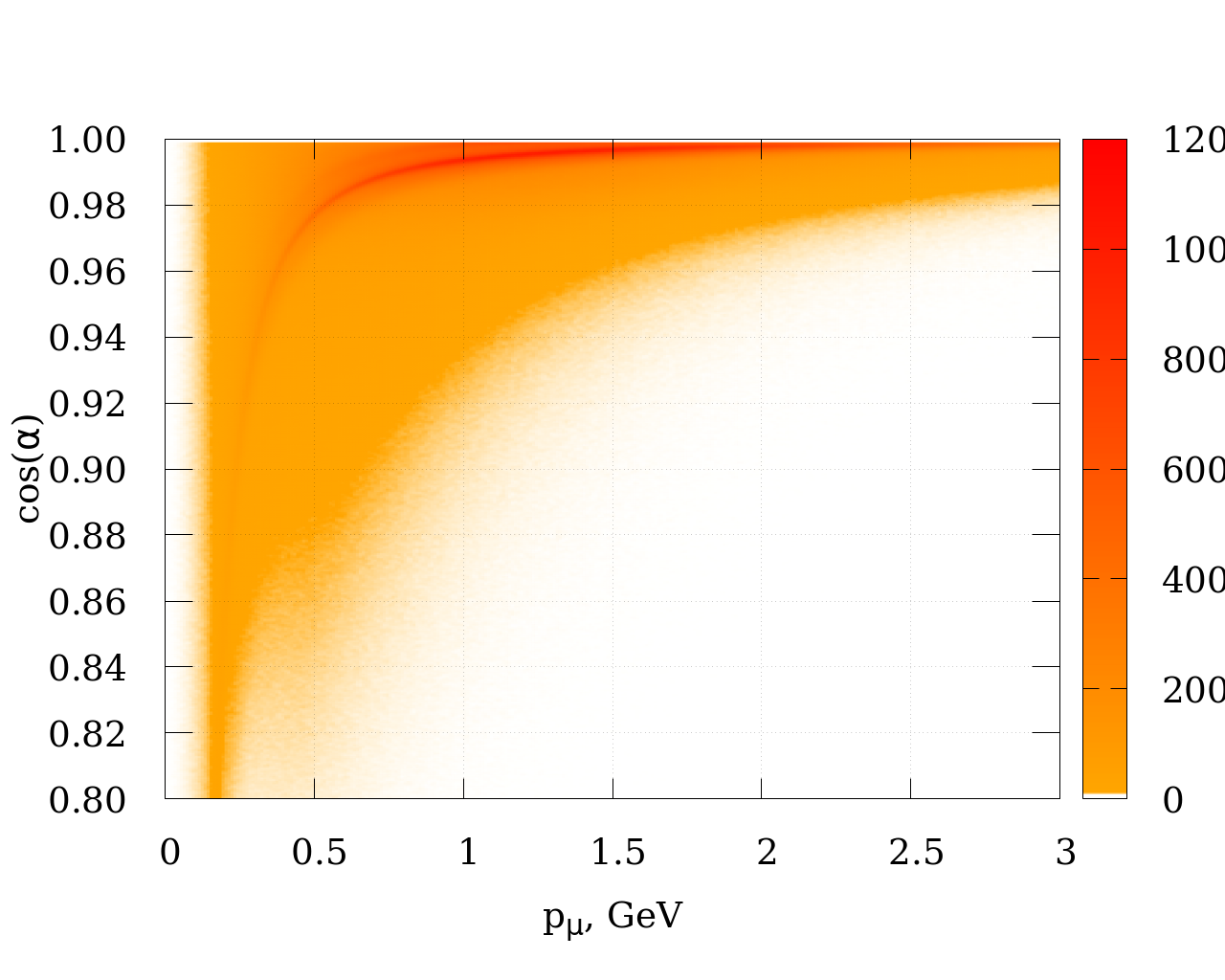} 
        \includegraphics[width=0.32\linewidth]{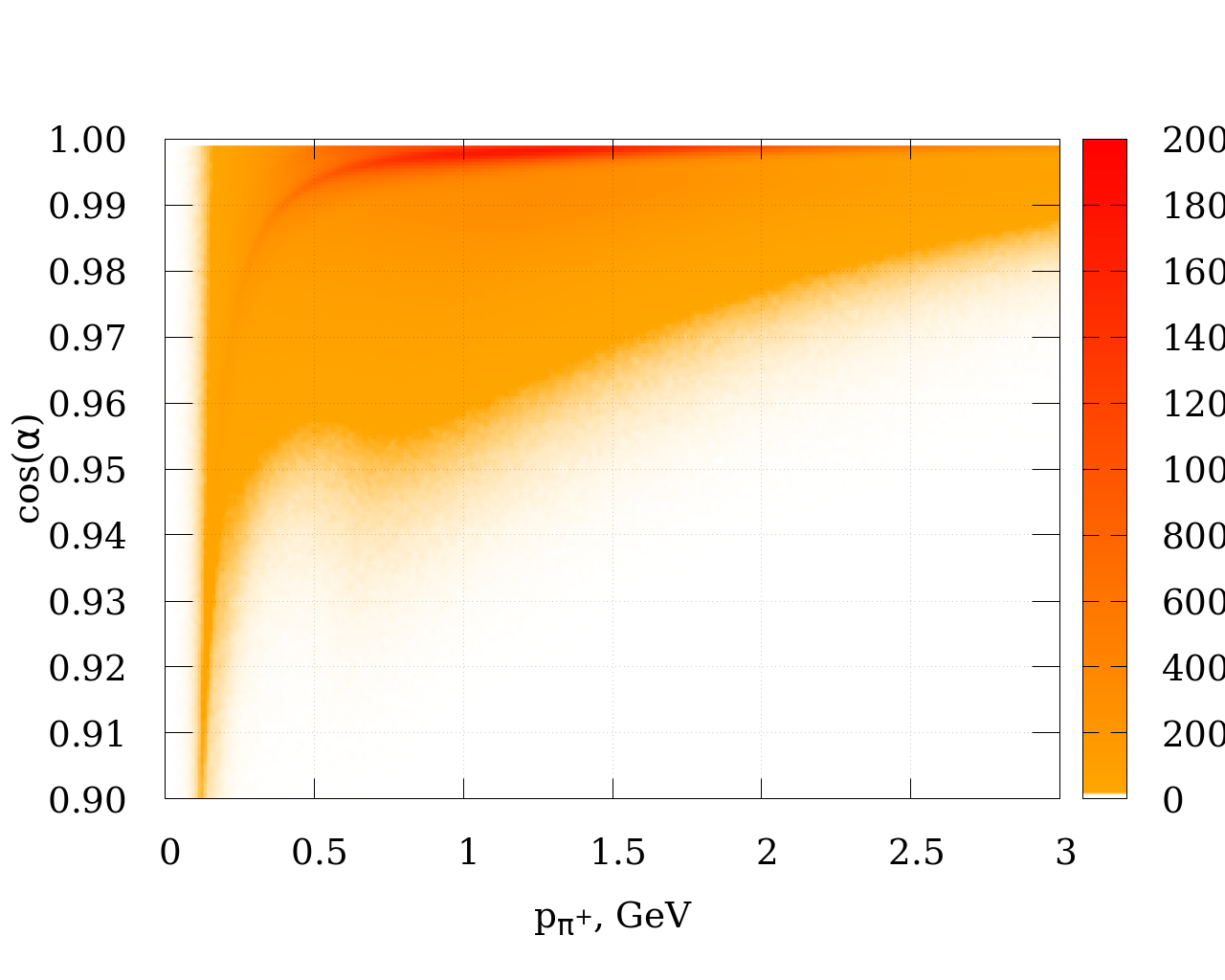} 
        }
    \caption{Kinematics of the daughter particles from scalar decay $K\to \pi \phi$, $\phi\to X^+X^-$: $K=K^\pm(K^0_L)$ on top (bottom) panels and $X^+X^-$=$e^+e^-$, $\mu^+\mu^-$, $\pi^+\pi^-$ from left to right, $M_\phi=300$ MeV.}
    \label{fig:products}
\end{figure}

To find 3-momenta of the light scalars we adopt the isotropic distribution in the rest frame of each decaying kaon and perform the Lorentz transformation back to the laboratory system with the help of a boost along the kaon 3-momentum.
We simulate the trajectories of the resulting light scalars and study further only those which pass through the detector. They form a fraction $\xi$ of all the produced light scalars, the so-called geometrical factor. 
Both the geometrical factor $\xi$ and the spectra of light scalars reaching the detector are shown in Fig. \ref{fig:xi+spectrum}. 
Naturally, the best chance to pass through the detector has scalars emerged in the decay tunnel. 
Using eqs.\,\eqref{eq:Gamma1},\eqref{eq:Gamma2} we calculate the probability of a light scalar of energy $E_\phi$ to decay in the detector volume:
\begin{equation}
\label{eq:P}
P = \left(1-\exp\left(-\Gamma_{tot} \;\Delta l\;M_\phi/p_\phi\right)\right)\; \exp\left(- \Gamma_{tot} \;d \;M_\phi/p_\phi\right), \quad\quad \Gamma_{tot} = \sum\Gamma(\phi\to...)\,.\nonumber
\end{equation}
The detection trigger in the PS191 experiment required at least two particles to reach the hodoscope~\cite{Bernardi:1987ek}. We applied the same condition to the scalar decay products.
To achieve this we simulated the decay and the trajectories of the daughter particles in the same way we simulated light scalars.
We counted as detectable only those decays for which both of the daughter particles pass through the backside of the detector.
The kinematic distributions of the daughter particles are shown in Fig.~\ref{fig:products}.
Bringing together all these factors we obtain the final estimate for the number of the light scalars that visibly decay in the detector volume
\begin{equation}
\label{Nsignal}
N_{\phi} = N_{POT} \times\sum_{K=K^+,\,K^-,\,K_L}\frac{N_{K}}{N_{sim}}  \cdot  Br(K \to \pi \phi) \cdot  \xi_K  \cdot P\,.
\end{equation}
Recall, that only the decay modes with charged particles can be recognized here, but it does not limit the PS191 power to test the light scalar, since the charged modes dominate, see Fig.\,\ref{fig:branchings+lifetime}, for the whole interval we are interested in. 

\begin{figure}[!htb]
    \centering
    \includegraphics[width=0.45\textwidth]{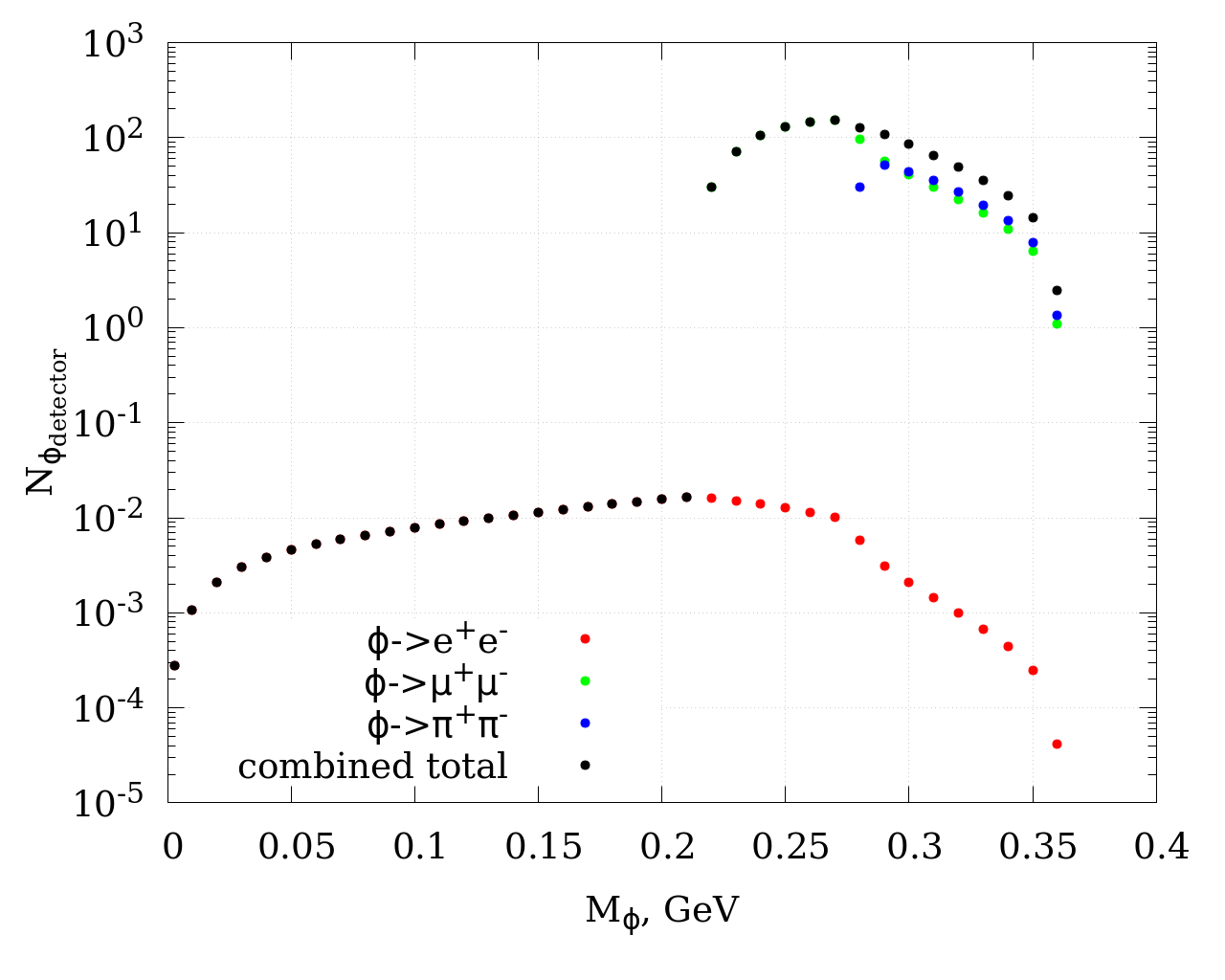}
    \includegraphics[width=0.45\textwidth]{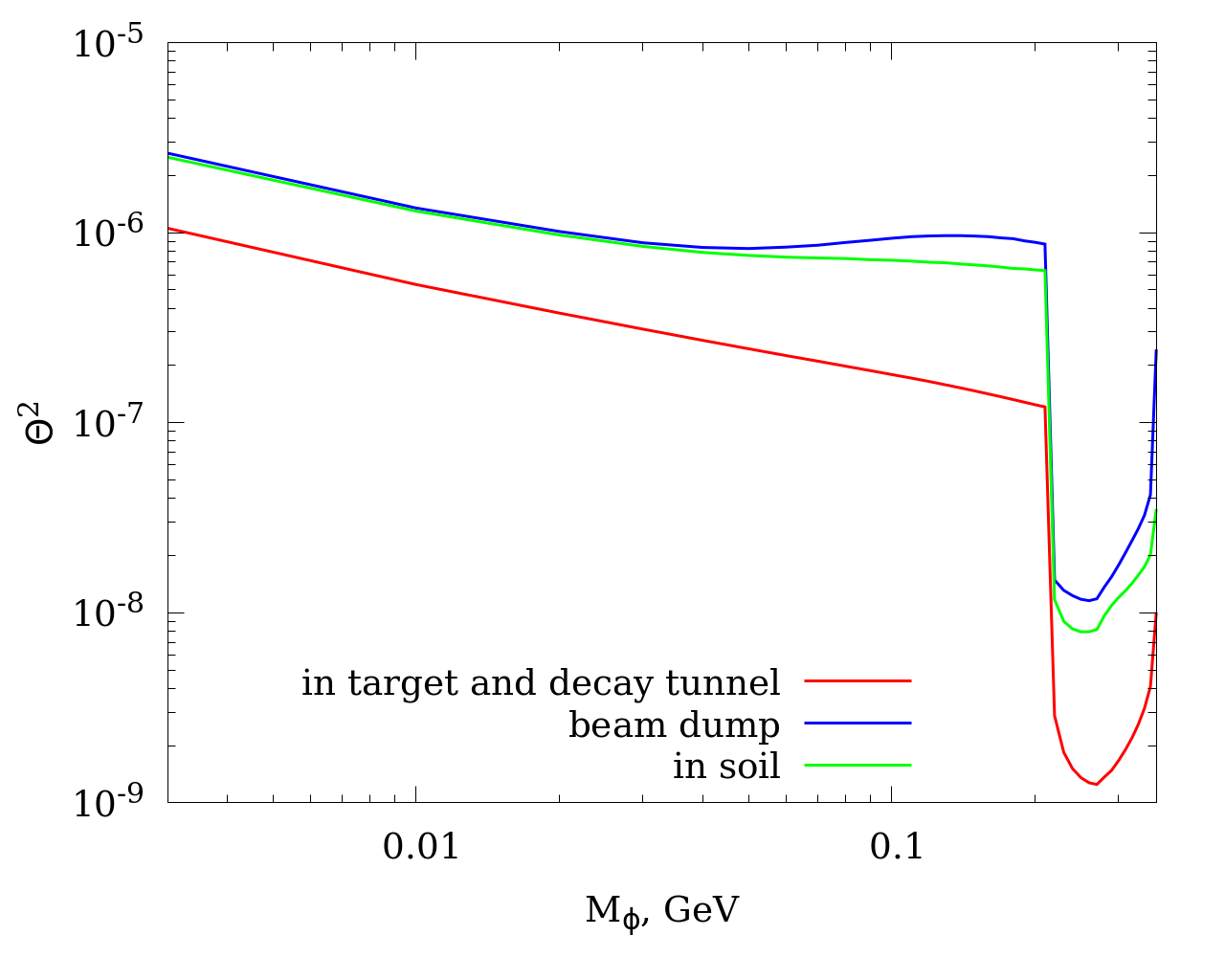}
    \caption{{\it Left panel:} the simulated number of light scalar decays in the detector volume in various decay modes, $\theta^2=10^{-8}$. {\it Right panel:} 
    Contribution of the different initial kaon decay positions to the final exclusion: the region above the solid line is excluded at 90\% CL from negative searches at PS191, $N_{POT}=0.3\times10^{19}$. 
    }
    \label{fig:N_phi+N_detector}
\end{figure}

Note that if the values of mixing are substantially big, the light scalars would decay before they reach the detector.
Therefore PS191 is unable to restrain area above some values of $\theta$.
We establish such a critical value of $\theta_{crit}$ as one that corresponds to a situation when the light scalar's mean path becomes equal to the distance from its production point to the detector: $\frac{p_\phi}{M_\phi} \frac{1}{\Gamma_{tot}(\theta_{crit})} = d$. From Fig. \ref{fig:N_phi+N_detector} one can discern that these values become relevant for the mass region $M_\phi>210$ MeV.

\begin{figure}[!htb]
    \centering
    \includegraphics[width=0.8\textwidth]{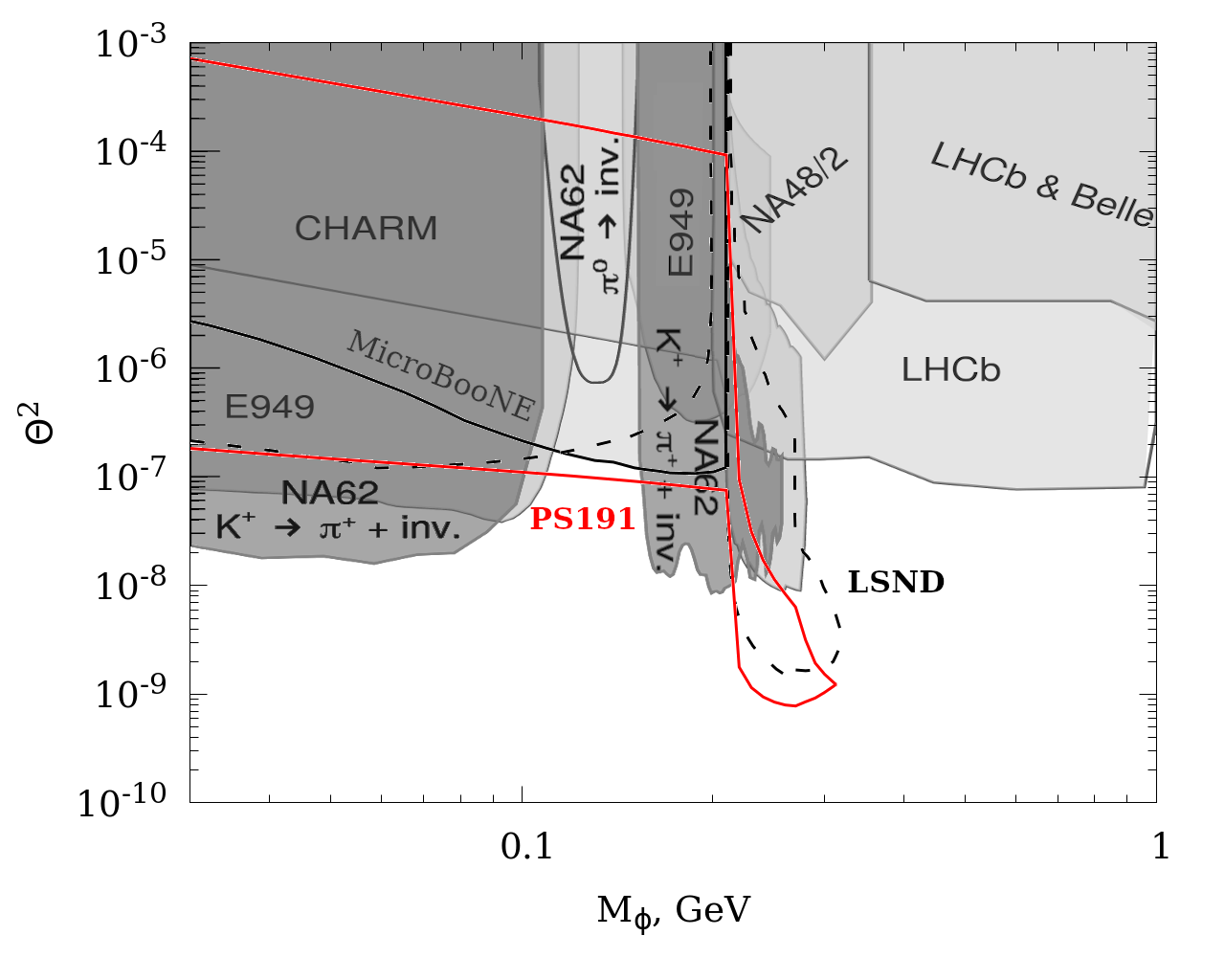}
    \caption{The model parameter space probed by various experiments: region outlined by the solid red line is excluded at 90\% CL from negative searches at PS191.
    }
    \label{fig:results}
\end{figure}

The final number of the light scalars that decay in the detector volume \eqref{Nsignal} is shown on the left panel of  Fig.\,\ref{fig:N_phi+N_detector}. 
And for a sufficiently small $\theta$ this number of the signal events scales as $N_{\phi_{detector}}\propto \theta^4$. We estimate the PS191 sensitivity by requiring for that number to be smaller than 2.3  (no signal at 90$\%$ CL): on the right panel of  Fig.\,\ref{fig:N_phi+N_detector} we show contributions of the different initial kaon decay positions to the final exclusion. 
Since no signal events were detected in PS191 experiment for all the charged modes, we sum all the contributions above, and so the region outlined by the red solid line in Fig.\,\ref{fig:results} gets excluded.  We adopted this plot from Ref.\,\cite{CortinaGil:2021nts} and added there also the recent reanalysis \cite{Foroughi-Abari:2020gju} of the LSND data and results of searches at MicroBooNE\,\cite{MicroBooNE:2021ewq} issued while our paper was under review.     

{\bf 6.} 
To summarize, we analyze the published results of PS191 experiment to place new constraints on models with light scalar mixed with the SM Higgs boson. In the parameter space of mixing angle and scalar mass, $(\theta,M_\phi)$ our study closes previously viable regions of masses 100-150\,MeV and $\theta^2\sim 10^{-7}\!-3\times\!10^{-7}$, and $M_\phi\sim 200\!-\!300$\,MeV, $\theta^2\sim 10^{-9}\!-\!10^{-8}$, see Fig.\,\ref{fig:results}.

Our results may be further refined. 
In this study, we neglect contributions of $K_S^0$, which decay branching ratio to a scalar is strongly suppressed. Likewise, the kaon scatterings off matter convert nucleons to hyperons, which can decay, producing the scalars. We ignore this source. We also do not perform simulations with magnetic fields part-time operated during PS191 runs, arguing that the limits based on GEANT4 simulations without the magnetic fields are conservative. We roughly approximated the soil around the decay tunnel and dump as sand, a substance of higher density would imply a larger geometrical factor and more signal events. However, the contribution of the 'in soil' region to the final limits is rather moderate, see right panel of Fig.\ref{fig:N_phi+N_detector}.   
It shows that the lack of details in the description of the 'in soil' region in our simulations can not change our results, which are conservative also in this respect.   

\vskip 0.3cm
We thank Yury Kudenko and Inar Timiryasov for their valuable comments. The work on the light scalar phenomenology is supported by the Russian Science Foundation  RSF grant 21-12-00379.

\bibliographystyle{utphys}
\bibliography{refs}
\end{document}